\documentclass[twocolumn]{aa}
\usepackage{graphicx,amssymb}
\newcommand \msun {\mbox{$\mathcal{M}_{\odot}$}}
\newcommand \kms {\mbox{km\,s$^{-1}$}}
\newcommand \degree {\mbox{$^\circ$}}

\begin{document}
   \title{Molecular Gas in NUclei of GAlaxies (NUGA) \\ III. The
     warped LINER NGC~3718 \thanks{Based on observations carried out
     with the IRAM Plateau de Bure Interferometer. IRAM is supported
     by INSU/CNRS (France), MPG (Germany) and IGN (Spain).}  }

   \subtitle{}

   \author{M. Krips\inst{1,2},
           A. Eckart\inst{1},
           R. Neri\inst{2},
           J.U. Pott\inst{1,11},
           S. Leon\inst{3},
           F. Combes\inst{4},
           S. Garc\'{\i}a-Burillo\inst{5},
           L.K. Hunt\inst{6},
           A.J. Baker\inst{7},
           L.J. Tacconi\inst{8},
           P. Englmaier\inst{12},
           E. Schinnerer\inst{9} \and
           F. Boone\inst{10}
          }

   \offprints{M. Krips, krips@ph1.uni-koeln.de}

   \institute{Universit\"at zu K\"oln, I.Physikalisches Institut, 
              Z\"ulpicher Str. 77, 50937 K\"oln, Germany\\
              \email{krips@ph1.uni-koeln.de}
              \email{eckart@ph1.uni-koeln.de}
            \and
              Institut de Radio-Astronomie Millim\'etrique (IRAM), 
	      300, rue de la Piscine, 38406 St.Martin-d'H\`eres,
              France\\
	      \email{neri@iram.fr}
            \and 
              Instituto de Astrof\'{\i}sica de Andaluc\'{\i}a (CSIC), 
	      C/ Camino Bajo de Hu\'etor,24, Apartado 3004, 18080 Granada,
              Spain\\
	      \email{stephane@iaa.es}
            \and 
             Observatoire de Paris, LERMA, 61 Av. de l'Observatoire,
              75014 Paris, France\\
	      \email{francoise.combes@obspm.fr}
            \and 
              Observatorio Astron\'omico Nacional (OAN)-Observatorio 
	      de Madrid, Alfonso XII, 3, 28014 Madrid, Spain\\
	      \email{burillo@oan.es}
            \and 
              Istituto di Radioastronomia/CNR, Sez. Firenze, Largo
              Enrico Fermi, 5, 50125 Firenze, Italy\\
	      \email{hunt@arcetri.astro.it}
            \and 
	      Jansky Fellow, National Radio Astronomy Observatory
	      Department of Astronomy, University of Maryland, 
	      College Park, MD 20742-2421 \\
	     \email{ajb@astro.umd.edu}
            \and 
              Max-Planck-Institut f\"ur extraterrestrische Physik,
              Postfach 1312, 85741 Garching, Germany\\
	      \email{linda@mpe.mpg.de}
            \and 
              Max-Planck-Institut f\"ur Astronomie, K\"onigstuhl 17, 
	      69117 Heidelberg, Germany\\
	      \email{schinner@mpia-hd.mpg.de}
            \and
              Max Planck Institut f\"ur Radioastronomie,
	      Auf dem H\"ugel 69, 53121 Bonn, Germany\\
              \email{fboone@mpifr-bonn.mpg.de}
	    \and 
	      ESO, Karl-Schwarzschild-Str. 2, 85748 Garching,
              Germany\\
	      \email{jpott@eso.org}
	    \and
	      Astronomisches Institut, Universit\"at Basel,
              Venusstr. 7, 4102 Binningen, Switzerland\\
	      \email{ppe@astro.unibas.ch}
             }

   \date{}

   \abstract{We present the first interferometric observations of
   CO(1--0) and CO(2--1) line emission from the warped LINER
   \object{NGC~3718}, obtained with the IRAM Plateau de Bure
   Interferometer (PdBI). This L1.9 galaxy has a prominent dust lane
   and on kiloparsec scales, a strongly warped atomic and molecular
   gas disk.  The molecular gas is closely associated with the dust
   lane across the nucleus and its kinematic center is consistent with
   the millimeter continuum AGN. A comparison of our interferometric
   mosaic data, which fully cover the $\sim 9\,{\rm kpc}$ warped disk,
   with a previously obtained IRAM 30\,m single dish CO(1--0) map
   shows that the molecular gas distribution in the disk is heavily
   resolved by the PdBI map.  On the nucleus the interferometric maps
   alone contain less than one half of the single dish line flux, and
   the overall mosaic accounts for about a tenth of the total
   molecular gas mass of $\sim 2.4\times 10^8 \msun$. After applying a
   short-spacing correction with the IRAM 30\,m data to recover the
   missing extended flux, we find in total six main source components
   within the dust lane: one associated with the nucleus, four
   symmetrically positioned on either side at galactocentric distances
   of about 1.3~kpc and 4.0~kpc from the center, and a sixth on the
   western side at $\sim 3\,{\rm kpc}$ with only a very weak eastern
   counterpart. In the framework of a kinematic model using tilted
   rings, we interpret the five symmetric source components as
   locations of strong orbital crowding. We further find indications
   that the warp appears not only on kpc scales, but continues down to
   $\sim 250\,{\rm pc}$. Besides the sixth feature on the western
   side, the lower flux (a factor of $\sim 2$) of the eastern
   components compared to the western ones indicates an intrinsic
   large scale asymmetry in NGC~3718 that cannot be explained by the
   warp. Indications for a small scale asymmetry are also seen in the
   central 600\,pc. These asymmetries might be evidence for a tidal
   interaction with a companion galaxy ({\it large scales)} and gas
   accretion onto the nucleus ({\it small scales}). Our study of
   NGC~3718 is part of the NUclei of GAlaxies (NUGA) project that aims
   at investigating the different processes of gas accretion onto
   Active Galactic Nuclei (AGN).

   }

 \authorrunning{M.Krips}

   \maketitle
%

  \begin{figure}[!]
   \centering
    \resizebox{\hsize}{!}{\rotatebox{-90}{\includegraphics{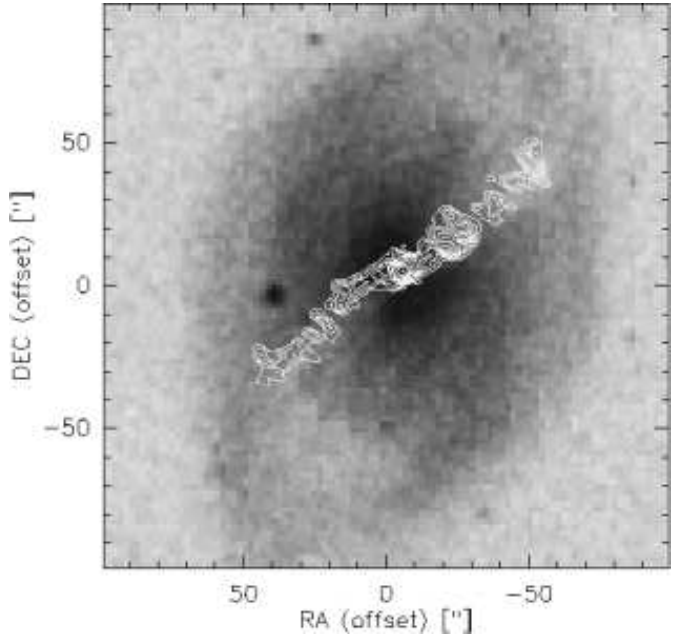}}}
      \caption{Optical image (taken from the DSS survey) superimposed
	with the integrated CO(1--0) contours (white) obtained at the
	PdBI, with added short spacings from the IRAM 30\,m telescope
	(see section~\ref{shsp}\,\&\,\ref{movsda}). Contour levels as
	in Fig.~\ref{shspall}.}
         \label{opt-pdbi30m}
   \end{figure}

\section{Introduction}

  \begin{figure}[t]
   \centering\resizebox{\hsize}{!}{\rotatebox{-90}{\includegraphics{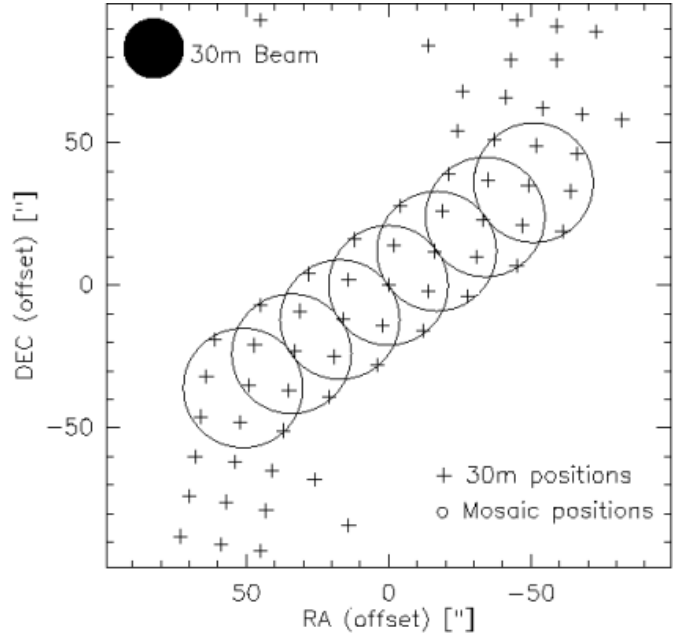}}}
      \caption{Positions of the IRAM 30\,m observations are plotted with
	       black crosses, and mosaic fields of the PdBI
	       observations are shown as circles indicating the 42$''$
	       primary beam size for each observing field. The 30\,m
	       beamsize is shown in the upper left.}
         \label{mospos}
   \end{figure}

Knowledge of the distribution and kinematics of the circumnuclear
molecular gas in active ga\-la\-xies is essential for understanding
the fueling of the central engine and the role of gas and dust in
obscuring the active nucleus (AGN). Although we have begun to
understand the factors that influence nuclear activity, a wealth of
unanswered questions still remain. In contrast to large scales ($\geq
3\,{\rm kpc}$) where dynamical perturbations like galaxy collisions,
mergers, or mass accretion (Heckman et al. 1986) as well as bars,
spirals and their gravity torques (Combes 1988) are responsible for
the infall of gas, the processes responsible for removing angular
momentum at small scales (sub-kpc) are not very well
understood. Various scenarios invoking nested bars (e.g., Shlosman et
al.  1993), spirals (e.g., Martini \& Pogge 1999), warped nuclear
disks (Schinnerer et al. 2000a,b), and lopsidedness or $m=1$
instabilities (Kormendy \& Bender 1999; Garc\'{\i}a-Burillo et
al. 2000) have been proposed. Millimeter interferometry allows
qualitative studies of molecular gas emission in these galaxies with
arcsecond spatial resolution (and even below), high spectral
resolution ($\sim$ a few \kms per channel) and high sensitivity
(detectability thresholds $\sigma_{\rm gas}\sim$ a few tens
\msun\,pc$^{-2}$). The NUclei of GAlaxies (NUGA) project
(Garc\'{\i}a-Burillo et al.\ 2003a,2003b) aims at establishing a
high-resolution and high-sensitivity CO survey of 12 nearby AGN
covering the full sequence of activity types including Seyfert 1 and
Seyfert 2 galaxies as well as Low Ionization Nuclear Emission-line
Region (LINER) and transition objects. The CO observations are
conducted with the IRAM Plateau de Bure Interferometer (PdBI:
Guilloteau et al.\ 1992), which currently provides the highest
sensitivity and resolution for the study of the distribution and
dynamics of molecular gas in the nuclei of these galaxies. Besides
case by case analyses and simulations of each object
(Garc\'{\i}a-Burillo et al.\ 2003b, Combes et al.\ 2004), the
collected data will also be used for a first-order approach to a
statistical study of how gas flows into nuclei and the different
mechanisms that account for further accretion inward.

One of the active galaxies belonging to the NUGA survey is NGC~3718
(Fig.~\ref{opt-pdbi30m}), a peculiar, elliptical galaxy at a distance
of 13\,Mpc (Pott et al.\ 2004; Schwarz 1985). Together with its
companion NGC~3729 it belongs to the Ursa Major group. It still
remains unclear if and to what extent a gravitational interaction is
taking place between these two galaxies. A large warped dust lane runs
across the entire stellar bulge of NGC~3718; it has a width of $<2''$
at the center and diverges into several smooth filaments across the
bulge ($1''$ corresponds to $\sim 64\,{\rm pc}$ for NGC~3718). At
a radius of about 1.5$'$ from the center, the dust lane bends by
almost 90 degrees towards the north and south. This warp signature is
also observed in the cold gas distribution, with the H\,I line
emission tracing it out to a radius of more than 6$'$ (24\,kpc) and CO
line emission tracing it towards the center ($\sim$ 20$''$: Pott et
al.\ 2004). The outer parts of the warp were kinematically modeled by
Schwarz (1985), while the inner parts down to $\sim 20'''$ were
modelled by Pott et al.\ (2004).


   \begin{table}
     \centering
     \begin{tabular}[c]{lccc}
           \hline
	   \hline
	          & \multicolumn {2}{c}{\hrulefill\hskip 2mm
    \raisebox{-0.8mm}{2000/01}\hskip 2mm \hrulefill}
                  & \multicolumn {1}{c}{\hrulefill\hskip 1mm
    \raisebox{-.8mm}{2001/02}\hskip 1mm \hrulefill} \\
                  & cont@  & cont@ & cont@ \\ 
	          & 3\,mm  & 1\,mm & 3\,mm  \\
           \hline
             R.A.   & 11:32:34.852 & 11:32:34.878 & 11:32:34.83 \\
           (hhmmss)                &  $\pm$0.003   & $\pm$0.005 & $\pm$0.01 \\
             Dec & 53:04:04.51  & 53:04:04.3 & 53:04:04.2 \\
	      (\degree:$':''$)                & $\pm$0.06 & $\pm$0.1 & $\pm$0.1   \\
             peak flux    &  9.4$\pm$0.7 & 10$\pm$2 & 7$\pm$1 \\ 
	      (${\rm mJy\,beam^{-1}}$) & & \\
	     \hline
	     \hline
     \end{tabular}
	 \caption[]{Interferometric positions and fluxes of the
	 continuum emission at 1\,mm and 3\,mm from the single field
	 data set (2000/01) and the central field of the mosaic
	 observations from 2001/02 ({\it right column}). The
	 difference in the 3\,mm flux densities between the first
	 epoch and the second is discussed in the text (section
	 \ref{continuum}).}
         \label{contab}
   \end{table}


  \begin{figure*}[!]
   \centering
    \resizebox{\hsize}{!}{\rotatebox{-90}{\includegraphics{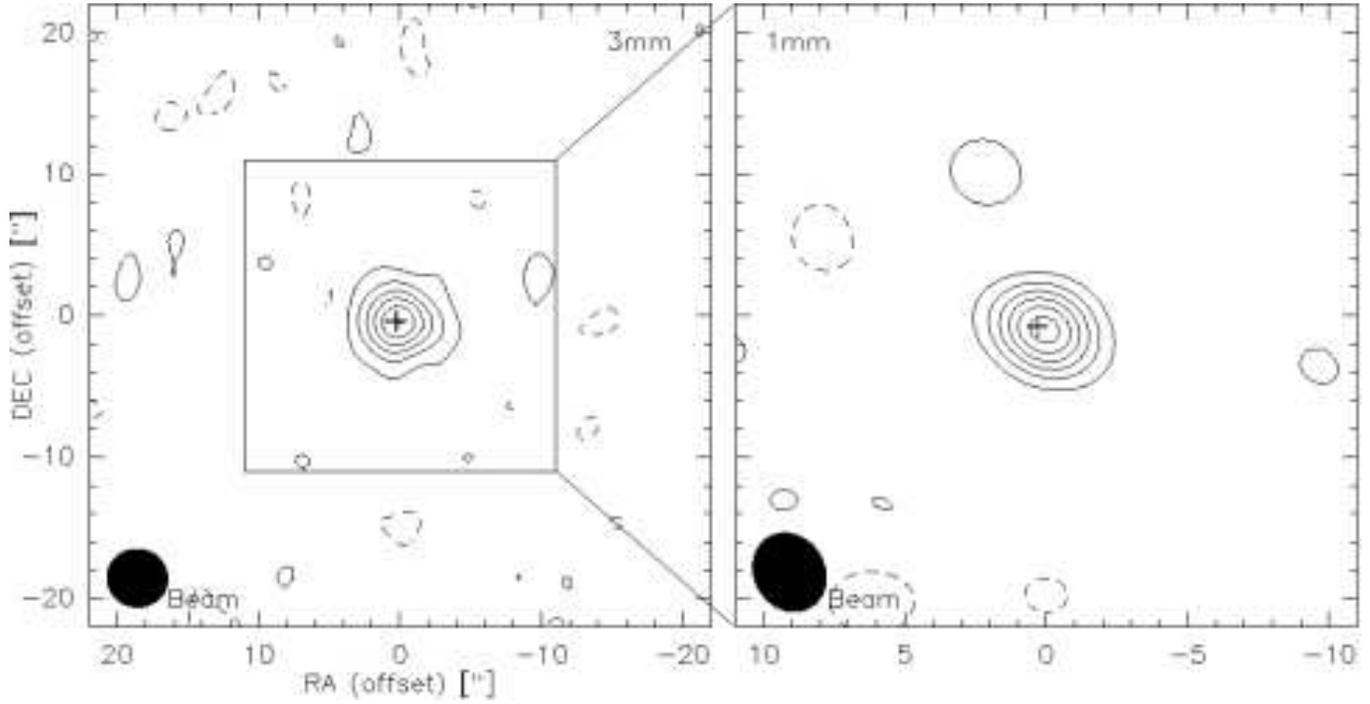}}}
      \caption{Continuum emission of the single field observations at
      3\,mm ({\it left}) and 1\,mm ({\it right}). The cross indicates
      the position at 18\,cm with EVN. Contour levels are from
      2$\sigma$ (2$\sigma$) = 1.0 (1.6)\,${\rm mJy\,beam^{-1}}$ to 9
      (9.6)\,${\rm mJy\,beam^{-1}}$ in steps of 4$\sigma$ (2$\sigma$)
      at 3\,mm (1\,mm). Beam sizes are indicated at lower left.}
         \label{pdbi-cen-con}
   \end{figure*}

  \begin{figure*}[!]
   \centering
    \resizebox{\hsize}{!}{\rotatebox{-90}{\includegraphics{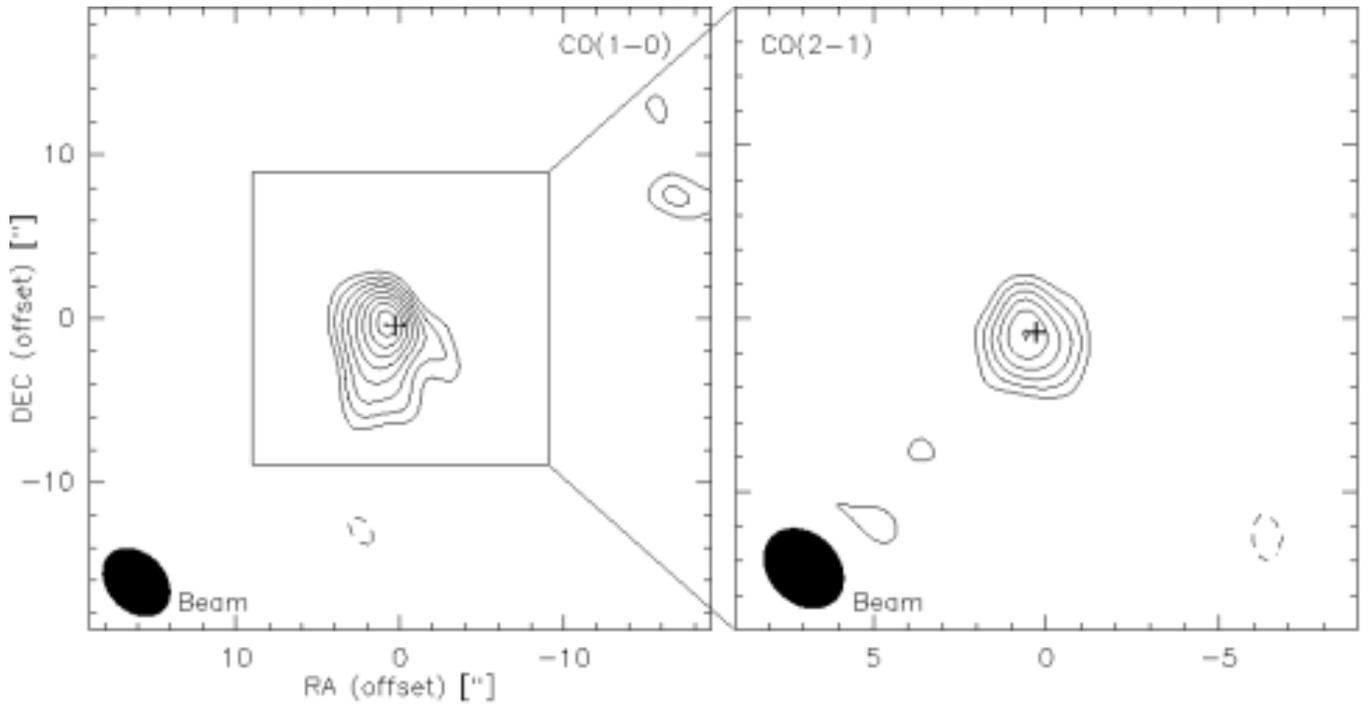}}}
      \caption{PdBI central maps of the velocity integrated CO(1--0)
	({\it left}) and CO(2--1) ({\it right}) emission in NGC~3718
	(from velocities $\sim-300$ to $\sim+300$\kms). Contour levels
	are from 3$\sigma$ (3$\sigma$) = 1.5 (2.7) to 5
	(7.2)\,Jy\,beam$^{-1}$\,\kms in steps of $1\sigma$ (1$\sigma$)
	for CO(1--0) (CO(2--1)).  The cross indicates the position at
	18\,cm with the European VLBI Network (EVN; Krips et al. 2005,
	in prep.).}
         \label{pdbi-cen}
   \end{figure*}

NGC~3718 is classified as a LINER galaxy of type 1.9 (Ho et al.\
1997). Weak, broad H$\alpha$ emission with FWHM $2350\,{\rm
  km\,s^{-1}}$ is emitted from the nucleus as well as strong [O\,I]
$\lambda$=6300\AA with FWHM $570\,{\rm km\,s^{-1}}$ indicative of a
hidden AGN (Filippenko et al.\ 1985; Ho et al.\ 1997). So far no
ultraviolet emission has been detected towards NGC~3718 (Barth et al.\
1998), probably due to obscuration of the nucleus by dust, whereas a
radio source has been confirmed at the position of the nucleus (this
work; Burke \& Miley 1973; Nagar et al.\ 2002).

In this paper we analyze single field and mosaic data obtained with
the PdBI at an angular resolution of $\sim 4''$. Section \ref{obs}
describes the observations. In section \ref{data} we present the
results from the observations. A kinematic model is outlined in
section \ref{sim}. We finish with a discussion.

\section{Observations}
\label{obs}

NGC~3718 was observed with the IRAM 30m and the IRAM PdBI. The IRAM
30m observations are already described in detail by Pott et al.\
(2004) and will thus here only be discussed in the context of the
short-spacing correction and resolution effects. The IRAM PdBI
observations were carried out in two different modes: {\it central
pointing} (one single field on the center) and {\it mosaic} (seven
fields along the dust lane). 

\subsection{Central pointing}
\label{centralfield}
During winter 2000/2001, we mapped the CO(1--0) and CO(2--1) lines in
a single field centered at the radio position of the AGN in NGC~3718,
i.e., $\alpha_{2000}=11^{\textrm h}32^{\textrm m} 34.840^{\textrm s}$
and $\delta_{2000}=53^{\circ} 04'04.90''$ (see
section\ref{continuum}). The PdBI was at this time deployed in the CD
set of configurations with 5 antennae. The bandpass calibrator was
3C273, while the phase and amplitude calibration were performed on
1150+497. Fluxes have been calibrated relative to CRL618 and
MWC349. The frequencies were centered on the redshifted
$^{12}$CO(1--0) line in the USB at 3\,mm and on the redshifted
$^{12}$CO(2--1) line in the LSB at 1\,mm. For each line, the total
bandwidth was 580\,MHz and the spectral resolution 1.25\,MHz. The
integration time for the central pointing amounts to $\sim 8$\,hr on
source. This gives a point source sensitivity of about $\sim$ 7\,mJy
($\sim$ 16\,mJy) in channels of 1.25\,MHz width at 3\,mm (1\,mm).  The
precipitable water vapor ranged between 4 and 10~mm (i.e., opacities
of $\sim$0.2-0.3) resulting in system temperatures of approximately
200-300~K on average.  We used the GILDAS software packages to reduce
and map the data (Guilloteau \& Lucas 2000). The synthesized beams are
4.7$'' \times 3.5''$ at 41\degree at 3\,mm and 2.6$'' \times 2.0''$ at
45\degree at 1\,mm.

\subsection{Mosaic}
\label{obsmosaic}
During winter 2001/2002, a mosaic of NGC~3718 was obtained with the
PdBI. Seven mosaic fields of the CO(1--0) emission were repeatedly
observed with six antennae or a subset in C and D configurations.
These sample the molecular gas emission along the dust lane with $\sim
12$\,kpc extent that was previously mapped with the IRAM 30\,m by Pott
et al.\ 2004 (Fig.~\ref{mospos}; see also Hartwich et al.\ 2002 and
Krips et al.\ 2003). The central field of the mosaic was set to the
same position specified in section \ref{centralfield}, and the
separation between the pointings was $21''$, corresponding to half the
primary beam size. The bandpass was calibrated on 3C84, 3C273, or
0923+392 with an accuracy of about 5\%. The latter source and
1150+497 were used as phase and amplitude calibrator, observed every
20 minutes. To derive the absolute fluxes of the calibrators and the
target source, values for each observing epoch were bootstrapped via a
standard IRAM flux monitoring program, yielding an accuracy of
the flux calibrations of $\sim 10\%$ at 3\,mm and $\sim 20\%$ at
1\,mm.

The spectral resolution was set to 1.25\,MHz at 114.89\,GHz, giving a
contiguous bandwidth of 580\,MHz. The integration time per field
amounted to $\sim 2.4$\,hr. The resulting point-source sensitivity is
$\sim$ 17\,mJy for a bandwidth of 1.25\,MHz and 5 antennae at
3\,mm. The precipitable water vapor ranged between 4 and 10~mm (i.e.,
opacities of $\sim$0.2-0.3) resulting in system temperatures of
approximately 200-300~K on average. Even if the CO(2--1) line had been
simultaneously observed in mosaic mode, the mosaic fields would only
have been usable as single fields and not as complete mosaic since the
separations of the mosaic fields are too large at 1\,mm where the
primary beam size is $\sim 21''$. Also, the sensitivity is too poor in
most of the fields, resulting in no detection of the CO(2--1) line
except in the central field.

  \begin{figure}[!]
   \centering
    \resizebox{\hsize}{!}{\rotatebox{-90}{\includegraphics{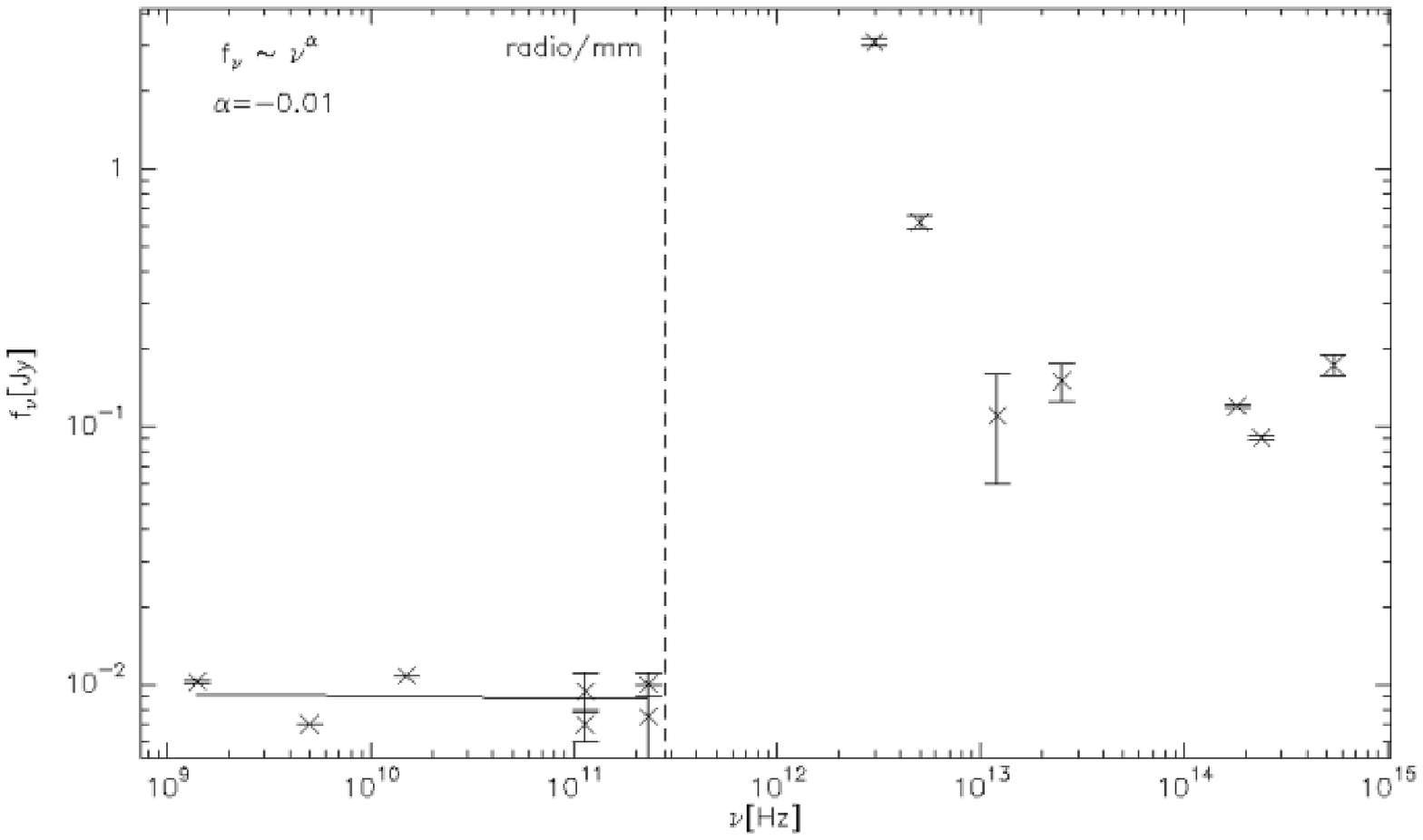}}}
      \caption{The Spectral Energy Distribution (SED) of NGC~3718. IR
	and optical fluxes with their errors are shown for
	completeness (taken from NED). The cm fluxes were observed
	with the VLA (FIRST: Becker et al.\ 1995; Nagar et al.\ 2001,
	2002) and thus have an angular resolution comparable to the
	PdBI maps. The mm fluxes for both observing epochs are plotted
	(upper limit at 1\,mm). The solid line indicates a fit
	(f$_\nu\propto\nu^\alpha$) with $\alpha=-0.01\pm0.1$.}
         \label{cont-si}
   \end{figure}

  \begin{figure}[!]
   \centering
    \resizebox{\hsize}{!}{\rotatebox{-90}{\includegraphics{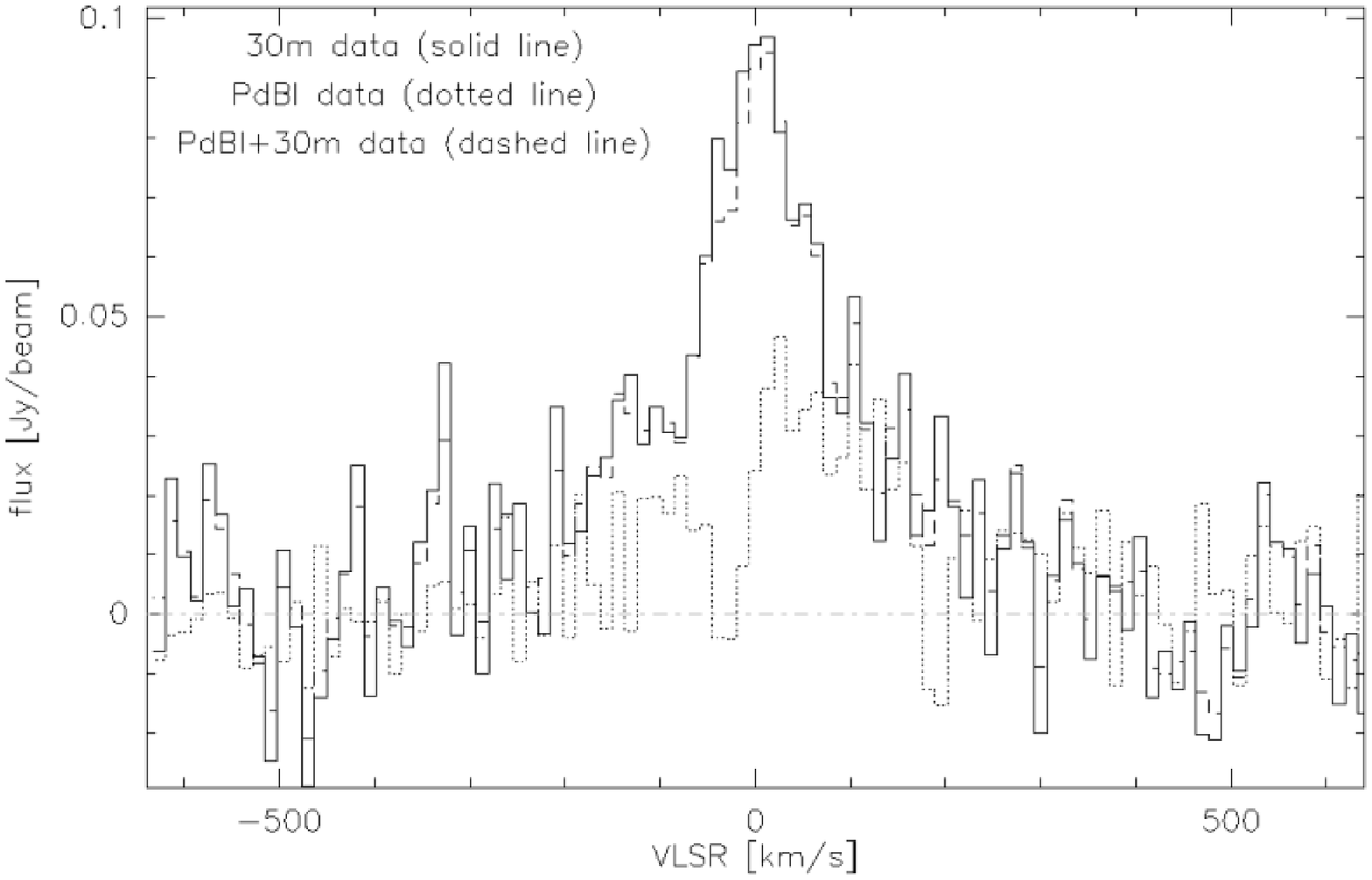}}}
    \resizebox{\hsize}{!}{\rotatebox{-90}{\includegraphics{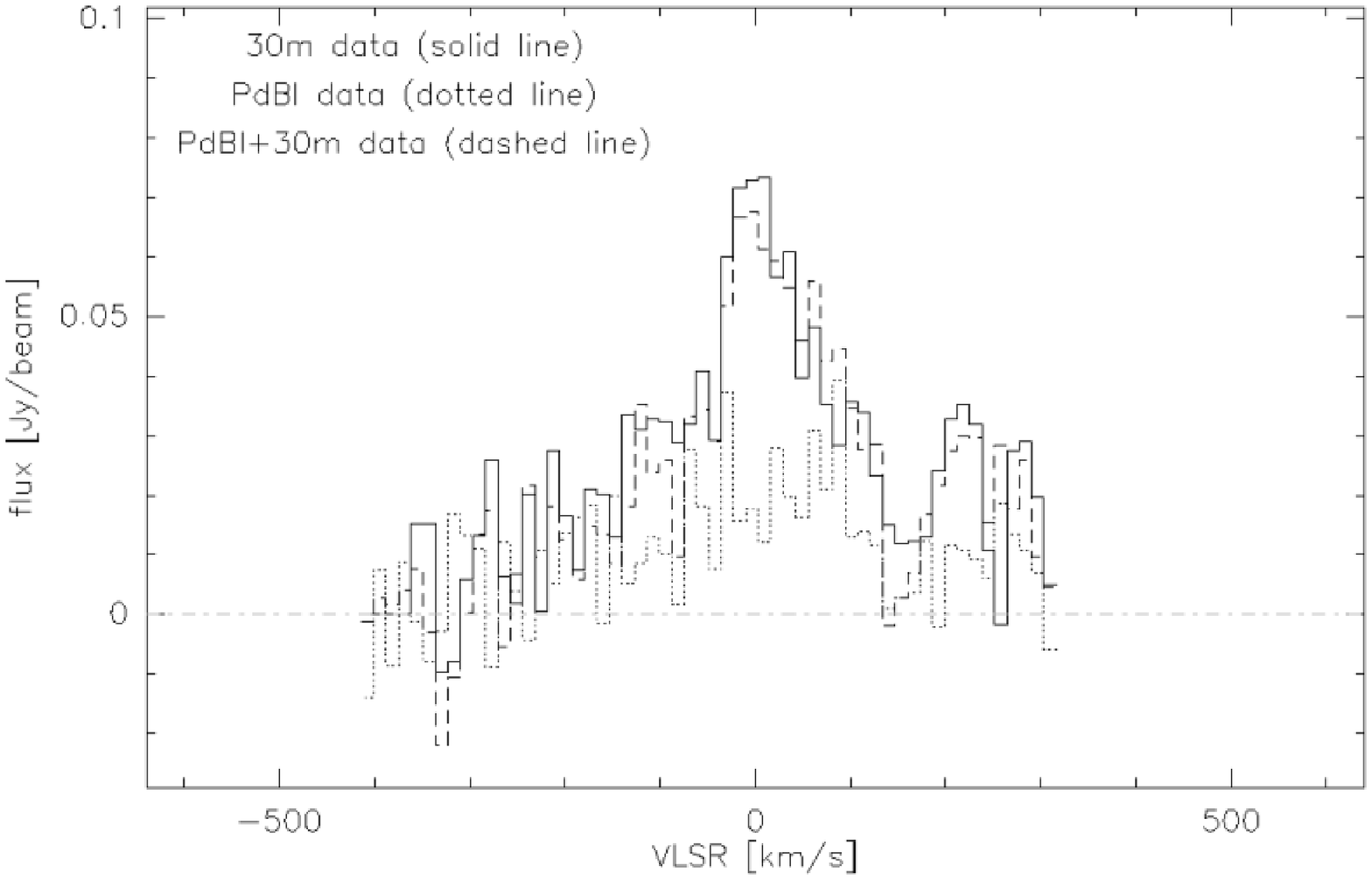}}}
      \caption{CO(1--0) ({\it upper panel}) and CO(2--1) ({\it upper
              panel}) spectra of the central emission in NGC~3718
              observed with the PdBI (dark grey; integrated over the
              emission area), the IRAM 30\,m telescope (light grey;
              Pott et al.\ 2004). The short-spacing corrected spectrum
              is also shown (dashed line; integrated over the emission
              area). Spectra are continuum subtracted. The rms noise
              level is here $\sim4$mJy ($\sim6$mJy) for the PdBI data
              for CO(1--0) (CO(2--1)) ($\Delta v$=13\kms for both).}
         \label{pdbi-cen-spec}
   \end{figure}

\section{The data}
\label{data}

\subsection{Central emission}
\label{central}
\subsubsection{Continuum}
\label{continuum}

Before combining the central pointing observation (single-field) with
the new central field from the mosaic observations, the continuum for
both data sets at 1\,mm and 3\,mm has been independently determined
from the line free channels. In both data sets at 3\,mm, i.e.\ single
field and central mosaic field, we detect faint continuum emission in
the central region of NGC~3718.  At 1\,mm, only the single-field data
set reveals continuum emission (see Fig.~\ref{pdbi-cen-con}); we do
not find any emission in the central mosaic field probably due to
sensitivity limitations for the 1~mm mosaic data.  At 3\,mm as well as
at 1\,mm, the continuum emission is point-like relative to the
respective beams (compare Table~\ref{contab}). The continuum declines
by $\sim 30\%$ between the two observing epochs at 3\,mm (see
Table~\ref{contab}). Assuming an upper limit of $3\sigma = 7.5\,{\rm
mJy\,beam^{-1}}$ at 1\,mm for the continuum of the central field from
the mosaic, we obtain a decline of $\geq 20\%$ relative to the single
field data (flux $\sim 10\,{\rm mJy\,beam^{-1}}$) consistent with the
3~mm data. As the atmospheric conditions were quite similar (similar
amounts of water vapor) between both data sets, we can mostly exclude
the possibility of atmospheric absorption which could cause an
artificial drop of the flux between the data. The phase and amplitude
calibrators were never separated from NGC3718 by more than a few
degrees in elevation.  The $\sim 10\%$ uncertainty in 3\,mm flux
calibration cannot explain all of the observed variability, although
this is not the case at 1\,mm. We do not find any support for a
variable continuum source at other wavelengths in recent EVN and
MERLIN measurements at 18\,cm and 6\,cm (Krips et al.\ in prep.,
epochs 2001 to 2004) or other cm observations (at 2\,cm and 3.6\,cm;
e.g., Nagar et al.\ 2002). In conclusion, the detected mm-variability
might be an intrinsic property of NGC~3718 but has to be further
investigated.

The derived positions are consistent between the two observing epochs
as well as between 1\,mm and 3\,mm. They also agree with the positions
of the nucleus obtained at cm and optical wavelengths within the
errors. Accounting for a weak variability, a comparison between our mm
fluxes and cm fluxes observed with the VLA in 1997 (Becker et al.\
1995), 1998 (Nagar et al. 2001), and 1999 (Nagar et al.\ 2002)
indicates a spectral index of -0.01$\pm$0.1 depending on which of our
two observing epochs is used. This corresponds to a flat to inverted
spectrum for the central source in NGC~3718, like those found for many
other low-luminosity active nuclei in Seyfert galaxies and LINERs
(Wrobel \& Heeschen 1991; Slee et al.\ 1994; Nagar et al.\ 2000). The
flat radio spectra exclude emission from optically thin synchrotron
emission, as is often seen from {\it extended} radio jets, since this
produces steep radio spectra. Thus, synchrotron self-absorption or
advection dominated accretion flows (ADAFs: Narayan, Mahadevan \&
Quataert 1998; Quataert 2001) have to be considered. A compact jet in
combination with an ADAF (Falcke 1996; Falcke \& Marloff 2000; Falcke
2001) also cannot be excluded. If a longitudinal pressure gradient
were to cause the jet to accelerate rapidly, the integrated radio
emission of the compact core jet would also have a slightly inverted
radio spectrum.

  \begin{figure*}[!]
   \centering
    \resizebox{\hsize}{!}{\rotatebox{-90}{\includegraphics{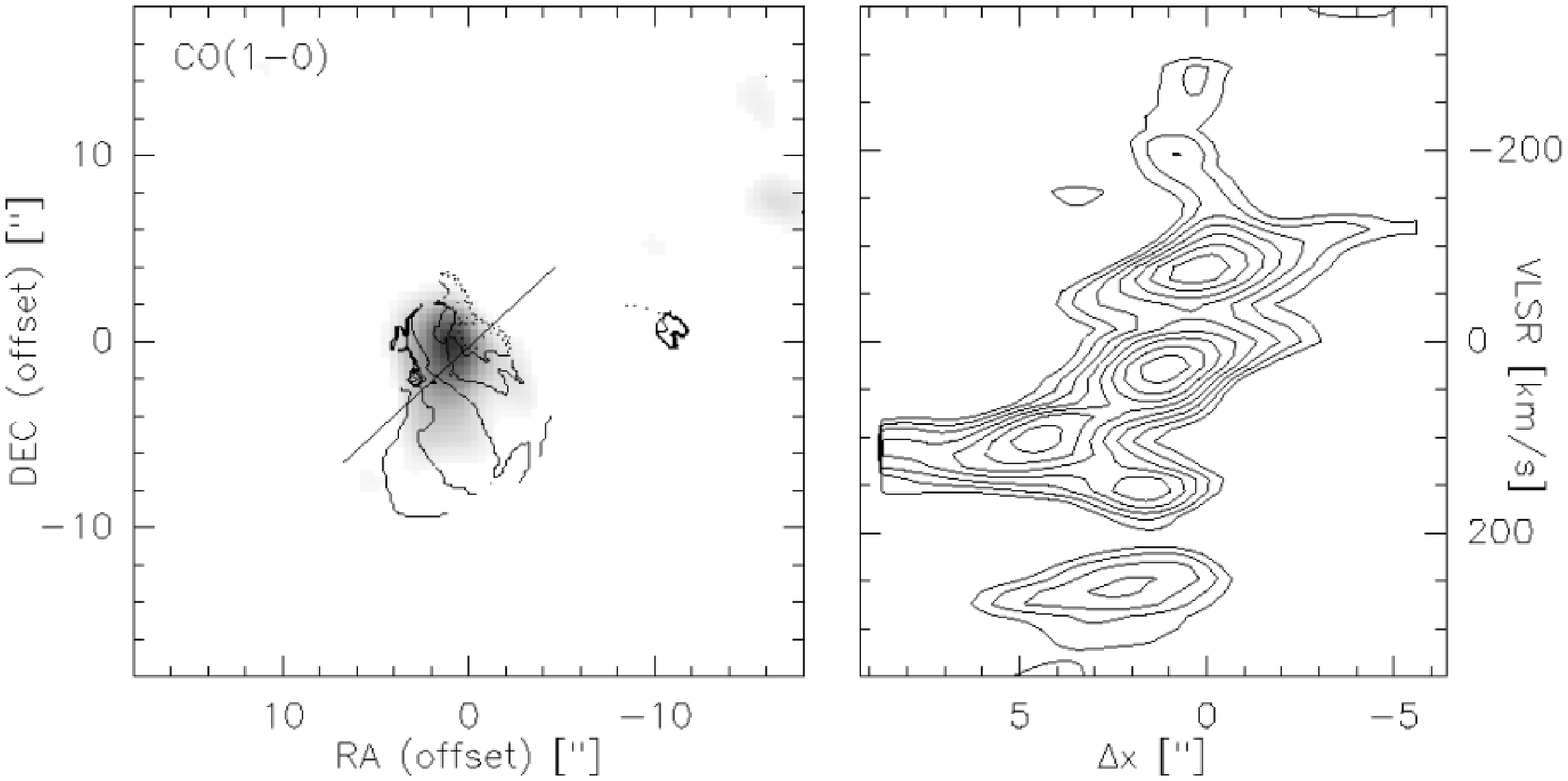}}}
    \resizebox{\hsize}{!}{\rotatebox{-90}{\includegraphics{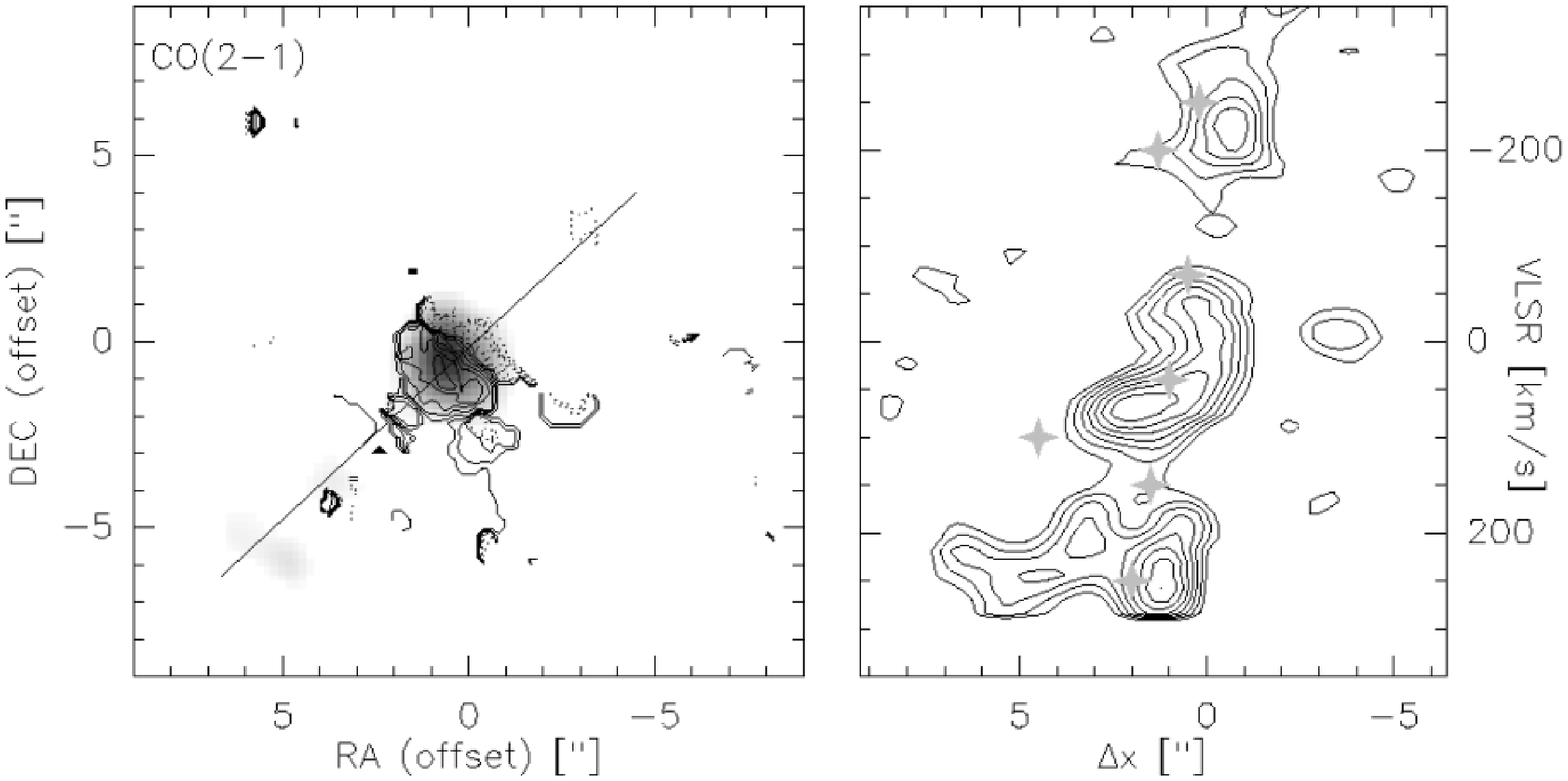}}}
      \caption{Mean velocity fields of the CO(1--0) ({\it upper left})
      and CO(2--1) ({\it lower left}) lines with their respective
      position-velocity maps (CO(1--0): {\it upper right}; CO(2--1)
      {\it lower right}) taken along the axis indicated in the maps
      ({\it solid line}: major axis). The grey scale in the images on
      the right is the integrated line emission of CO(1--0) and
      CO(2--1) respectively. The light grey stars in the lower right
      panel illustrate the positions of the CO(1--0) components
      identified in the upper right panel. {\it Left:} Velocity
      contours are in steps of 25\kms for CO(1--0) and for
      CO(2--1). {\it Right:} Contour levels are in steps of 10\% from
      the peak starting at 30\%. The position velocity maps were
      Hanning smoothed.}
         \label{pdbi-co10-cen-velo}
   \end{figure*}

  \begin{figure*}[!]
   \centering
    \resizebox{\hsize}{!}{\rotatebox{-90}{\includegraphics{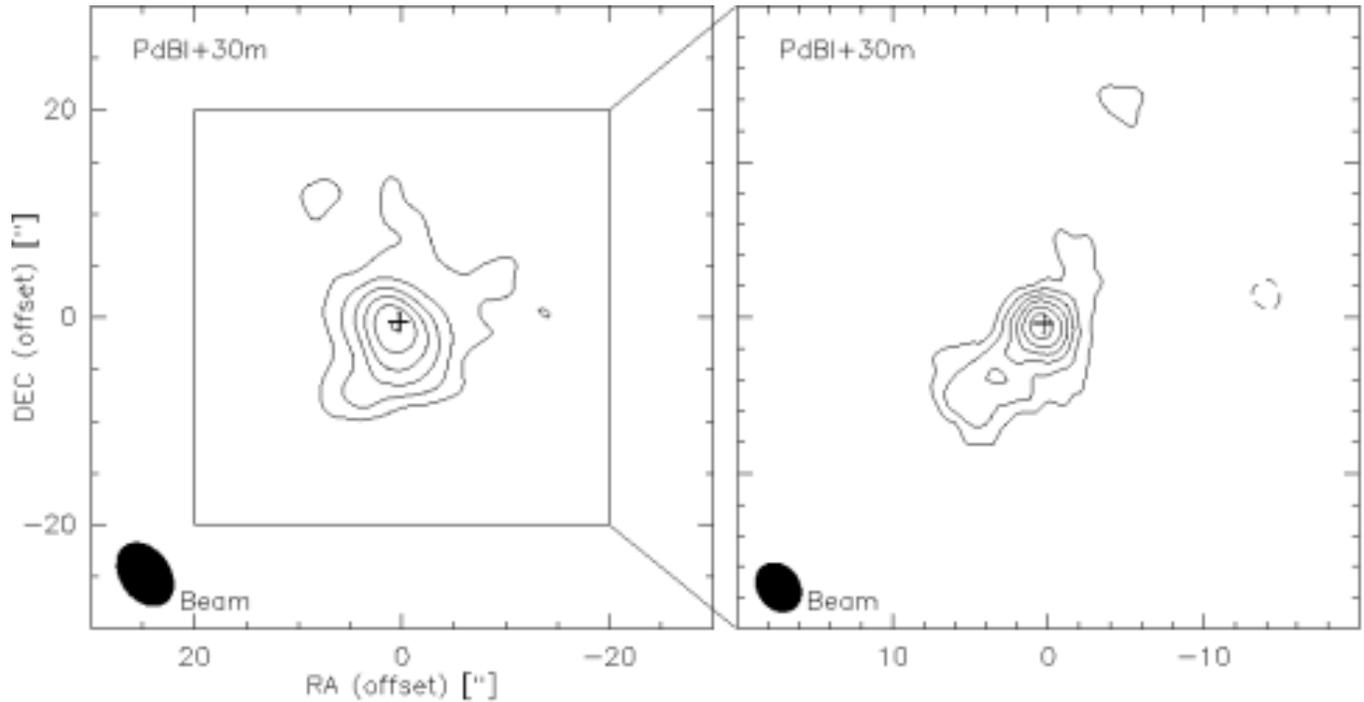}}}
      \caption{PdBI central maps corrected with 30m data for
	short-spacing effects of the integrated CO(1--0) ({\it left})
	and CO(2--1) ({\it right}) emission in NGC~3718 at same
	size. Contour levels are from 5$\sigma$ (3$\sigma$) = 2.0
	(2.4) to 8.0 (10.4)\,Jy\,beam$^{-1}$\,\kms in steps of
	$3\sigma$ (2$\sigma$) for CO(1--0) (CO(2--1)).  The white
	cross indicates the position of the AGN radio continuum source
	at 18\,cm derived from EVN data.}
         \label{pdbicen-shsp}
   \end{figure*}

  \begin{figure*}[!]
   \centering
    \resizebox{\hsize}{!}{\rotatebox{-90}{\includegraphics{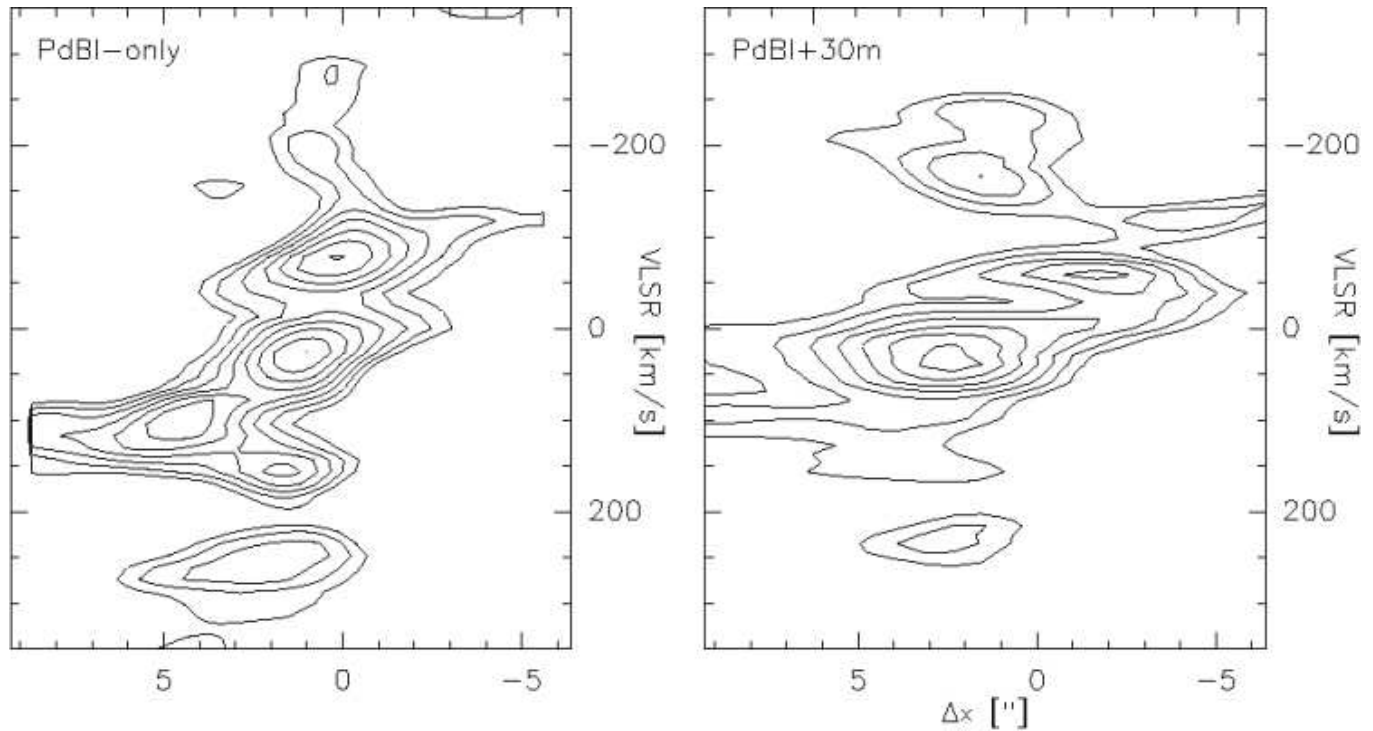}}}
      \caption{Comparison of pv-diagram along axis shown in
               Fig.~\ref{pdbi-co10-cen-velo} between PdBI-only data
               for the central pointing ({\it left}) and PdBI+30\,m
               data for the central mosaic ({\it right}). The
               asymmetry is lowered in the central mosaic map but does
               not disappear totally. Thus, it cannot be totally due
               to missing short spacings. Contour levels are in steps
               of 10\% from the peak starting at 30\%. The position
               velocity maps were Hanning smoothed.}
         \label{velo-pdbi-pdbi30m}
   \end{figure*}

  \begin{figure}[!]
   \centering
    \resizebox{\hsize}{!}{\rotatebox{-90}{\includegraphics{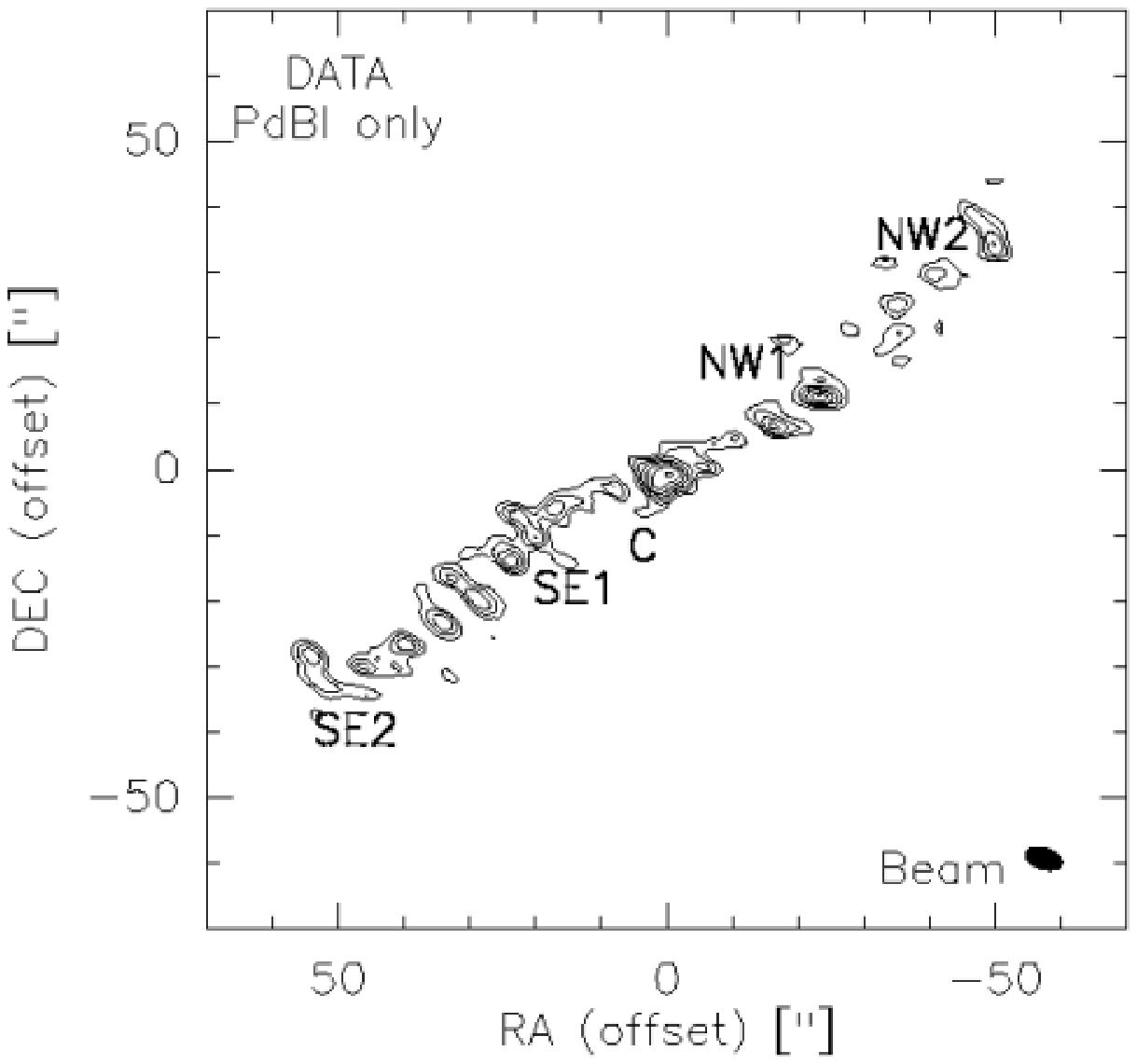}}}
     \vskip 0.2cm
     \hskip -0.2cm  
     \resizebox{7cm}{!}{\rotatebox{-90}{\includegraphics{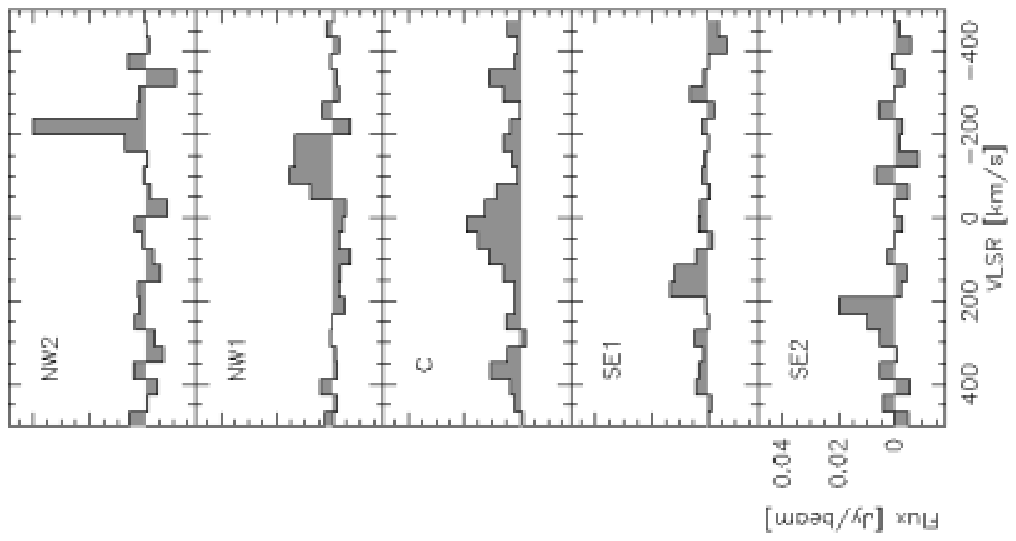}}}
      \caption{PdBI mosaic map of the velocity integrated CO(1--0)
	emission in NGC3718 (upper figure; from $-$300 to
	$+$300\kms). Spectra are shown in the lower panel. Contour
	levels are from $2\sigma=$1.0\,Jy\,beam$^{-1}$\,\kms to
	4\,Jy\,beam$^{-1}$\,\kms in steps of $1\sigma$.}
         \label{pdbi}
   \end{figure}

  \begin{figure}[!]
   \centering
    \resizebox{\hsize}{!}{\rotatebox{-90}{\includegraphics{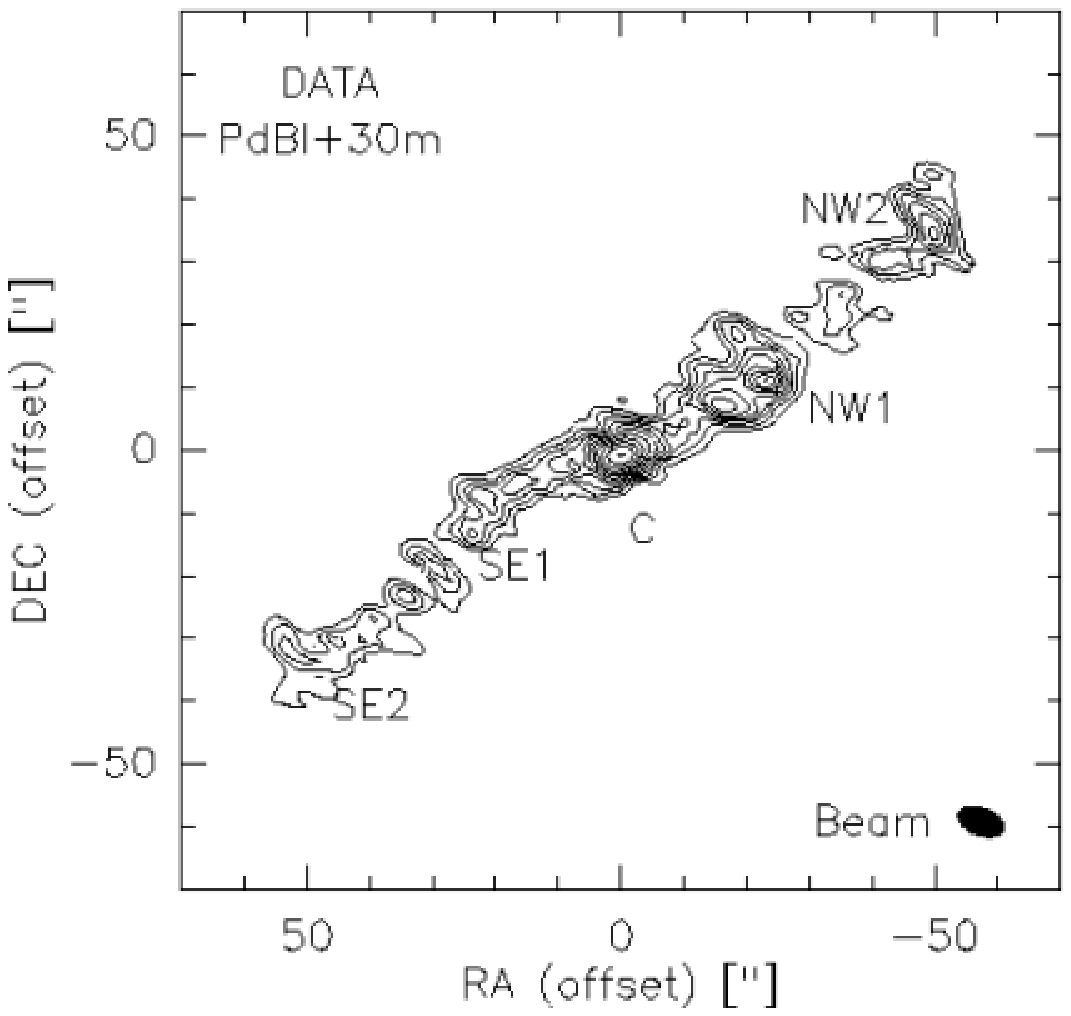}}}
     \vskip 0.2cm
     \hskip -0.2cm  
    \resizebox{7cm}{!}{\rotatebox{-90}{\includegraphics{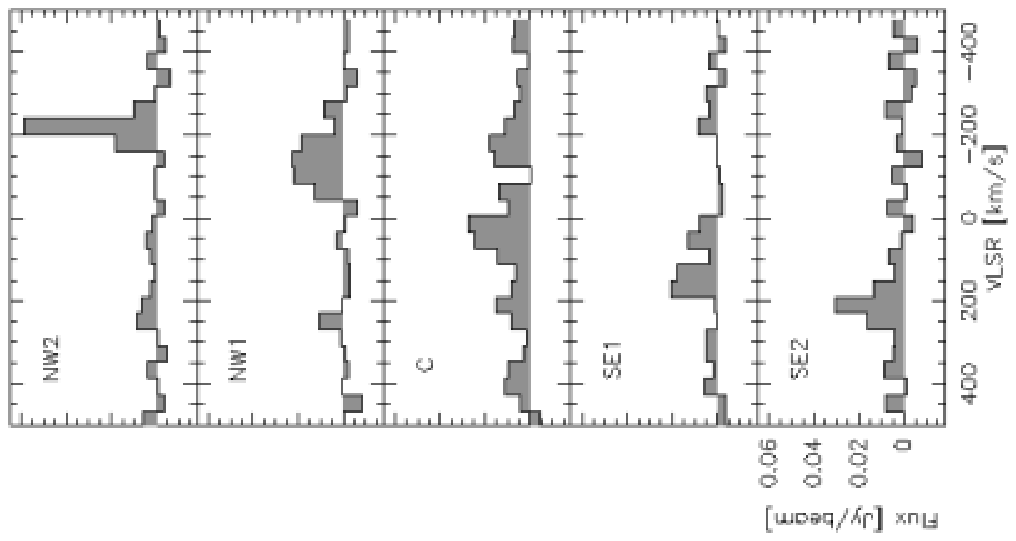}}}
      \caption{Velocity integrated intensity map of the CO(1--0)
      emission for the combined PdBI+30m data (fromm $-$300 to
      $+$300\kms). Contour levels are from
      $2\sigma=$1.0\,Jy\,beam$^{-1}$\,\kms to
      6.0\,Jy\,beam$^{-1}$\,\kms in steps of $1\sigma$.}
         \label{data-pdbi30m-r4}
   \end{figure}

\subsubsection{Line}
\label{line}

After subtracting the continuum emission in the $uv$ plane from the
(continuum+line) channels, CO(1--0) and CO(2--1) line emission is
detected in the central $\sim$ 10$''$ ($\equiv$ 1\,kpc) of
NGC~3718. The combined central maps are shown in
Fig.~\ref{pdbi-cen}. In both lines the emission is slightly
extended. The positions of the integrated CO(1--0) and CO(2--1)
emission are similar to the nuclear position derived from the
radio. Also, the iso-velocity diagram in Fig.~\ref{pdbi-co10-cen-velo}
shows that the dynamical center of the CO emission lies on the radio
nucleus. A comparison of the flux observed with the PdBI and the IRAM
30\,m single dish reveals that the interferometer recovers less than
one half of the peak line flux seen with the 30\,m telescope (see
Fig.~\ref{pdbi-cen-spec}).  This indicates that the emission in
NGC~3718 must be much more extended and diffuse than is apparent in
the integrated map of Fig.~\ref{pdbi-cen}.  Assuming a standard
$M_{\rm gas}$(=$M$(H$_2$+He)=1.36 $\times M$(H$_2$)) to $L'_{CO(1-0)}$
ratio of $\sim5$\msun(K\kms pc$^2$)$^{-1}$ (Solomon \& Barrett 1991),
the central molecular gas mass $M_{\rm gas}$ amounts to
$\sim1\times10^7$\msun in the PdBI map, in contrast to
$3\times10^7$\msun in the 30\,m map (Pott et al.\ 2004). Downes et
al.\ (1993) emphasize, however, that the standard galactic conversion
factor between gas mass and CO luminosity might fail for the centers
of galaxies where the molecular gas is bound by the total
gravitational potential of the galaxy rather than by self-gravity.  In
such a case, the conversion factor would be up to 5 times lower than
the standard galactic one, resulting in gas masses of $\sim 0.2 \times
10^7$\,\msun for the PdBI map and $0.6 \times 10^7$\,\msun for the
30\,m map.

By comparing the distributions of the central CO(1--0) and CO(2--1)
line emission, we find that the CO(1--0) emission shows an elongation
towards the south, almost perpendicular to the beam.  We estimate the
integrated line ratio $R_{21}=$CO(2--1)/CO(1--0) to $\sim0.6$
indicating subthermal excitation conditions and cold gas. To obtain
this value, we have first truncated the inner $uv$ region of the
CO(1--0) data and the outer $uv$ region of the CO(2--1) data to get
comparable $uv$ coverages and resolutions for the two transitions,
integrated both lines over the same velocity range of -300\kms to
+300\kms and then transformed the flux (Jansky) into temperature
(Kelvin). However, due to the limited signal-to-noise and the
difference of the weighting factors between both transitions, the
reliability of this estimate is still questionable and thus the value
of $\sim0.75$ derived in the center by Pott et al.\ (2004) might be a
better indicator for the physical conditions in NGC~3718.

Fig.~\ref{pdbi-co10-cen-velo} shows the zero order moment maps of both
transitions overlaid with contours from the first order moment maps
and the position velocity cuts taken along the kinematic major
axis. The CO(2--1) maps are quite noisy but are plotted for
consistency and comparison purposes. The position velocity diagram of
the CO(1--0) line emission reveals a quite extended ($\sim8''$)
velocity gradient between $\pm150$\kms. Two further features appear at
very high velocities ($\sim\pm220$\kms) with a somewhat steeper
gradient ($\sim\pm2''$).  Consistent with Pott et al.\ (2004), the
global velocity range of the inner $\sim 15''$ amounts to almost
$\pm$220\kms. However, this ``central'' feature in the position
velocity diagram extends in Pott et al. (2004) out to a radius of
15-20$''$, while in our interferometric data it stops at a radius of
$\lesssim10''$. This rapidly rotating nuclear component will be
discussed in section~\ref{movsda}. The line emission shows a strong
asymmetry in the position velocity diagram with only little emission
towards the blueshifted velocities. Section~\ref{datashsp} will show
that this asymmetry cannot be totally traced back to missing short
spacings.

\begin{table}
\centering
  \begin{tabular}[c]{cccc}
     \hline
     \hline
     Component & R.A. (offset) & Dec. (offset) & Flux \\
     & [$''$]        & [$''$]        & Jy\kms \\
     \hline
     SE2        & +52$\pm$3 & -34$\pm$3  & 1.2\\
     SE1        & +20$\pm$3 &  -8$\pm$3  & 1.3\\
     C         &   0$\pm$1 &   0$\pm$1  & 3.4\\
     NW1        & -19$\pm$1 &  +9$\pm$1  & 2.9\\
     NW2        & -49$\pm$1 & +36$\pm$1  & 2.7\\
     \hline
\end{tabular}
\caption{Positions and fluxes for CO(1--0) components in the PdBI+30m
map. Offsets are with respect to $\alpha_{2000}$=11h32m34.8 and
$\delta_{2000}$=53$^\circ$04$'$04.9$''$. }
\label{pos-pdbi30m}
\end{table}

\subsection{Overall emission}
\label{mosaic}

\subsection{PdBI-only data}
\begin{figure}[!]
   \centering
    \resizebox{8.5cm}{!}{\rotatebox{-90}{\includegraphics{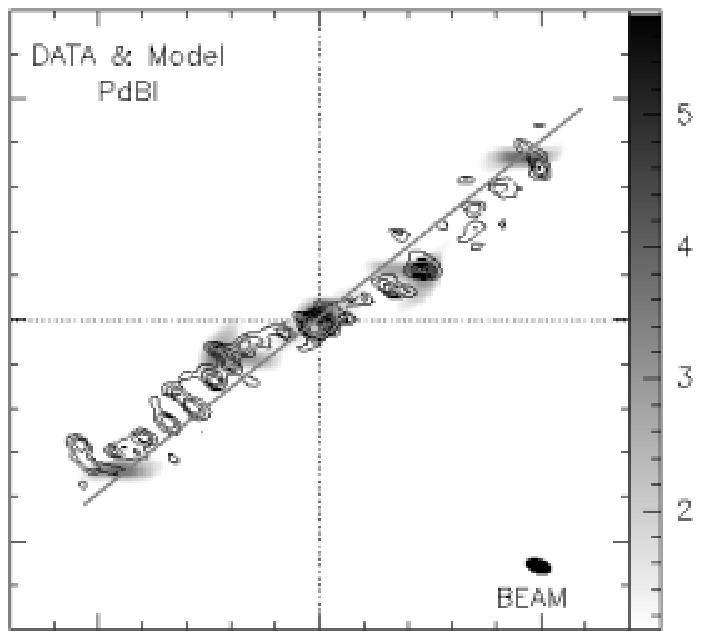}}}
    \resizebox{8.5cm}{!}{\rotatebox{-90}{\includegraphics{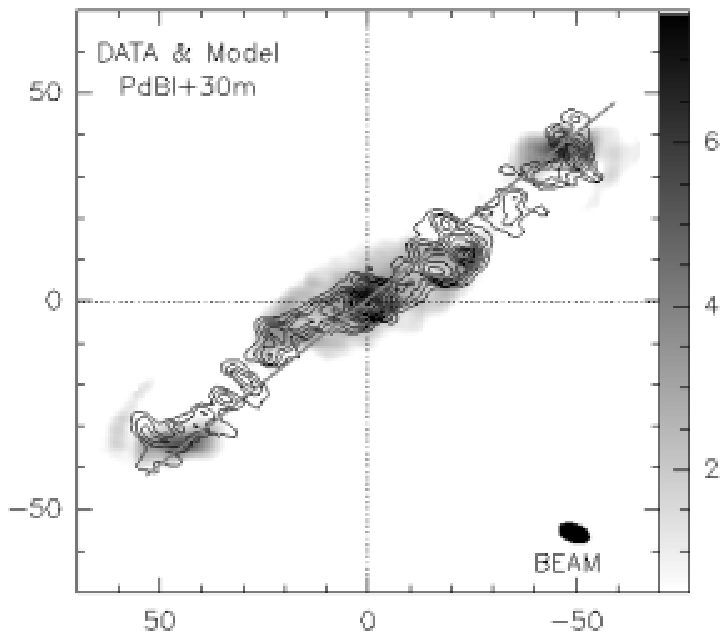}}}
    \resizebox{8.5cm}{!}{\rotatebox{-90}{\includegraphics{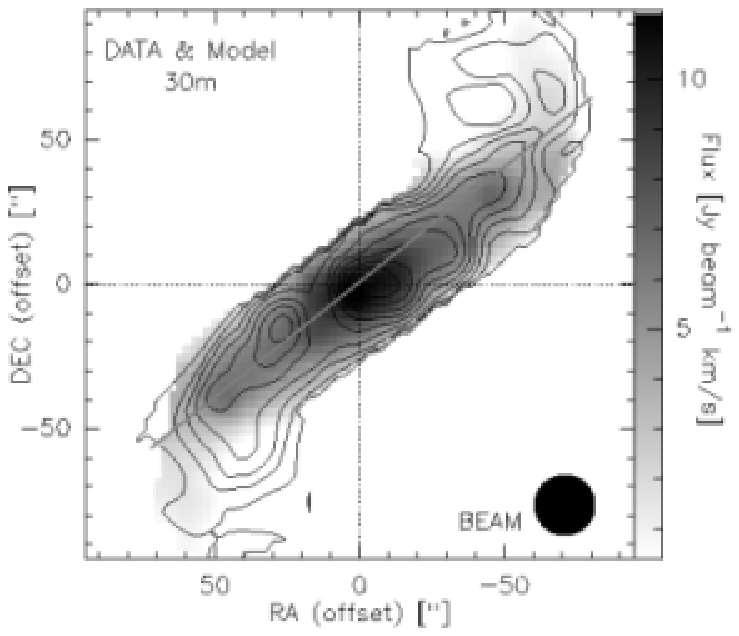}}}
      \caption{Model vs. observation. Integrated maps of the PdBI data
      only ({\it upper panel}), the PdBI and 30\,m data combined ({\it
      middle panel}) and the 30\,m data only ({\it lower panel}). The
      model (grey scale) is overlaid with solid contours of the
      observed data. Contour levels: {\it PdBI-only} $2\sigma$=1.0 to
      4\,Jy\,beam$^{-1}$\,\kms by $1\sigma$; {\it PdBI+30m}
      $2\sigma$=1.0 to 6.0 \,Jy\,beam$^{-1}$\,\kms by $1\sigma$; {\it
      30m-only} $3\sigma=$1.0 to 10 by 1.0\,Jy\,beam$^{-1}$\,\kms.}
         \label{shspall}
   \end{figure}

Outside the central mosaic field, there is no evidence of continuum
emission in our data. The 30\,m map of Pott et al.\ (2004) has already
revealed the existence of two additional maxima in the CO(1--0)
emission which are offset by $\sim 30-40''$ from the nucleus. We find
further maxima in the new PdBI mosaic map which are located farther
out at $\sim\pm60''$ (compare Figs. \ref{pdbi} and \ref{shspall}).  In
Fig.~\ref{pdbi} we have labeled the eastern components SE1 and SE2,
the western components NW1 and NW2, and the central feature C.  There
are also indications for further maxima between NW1 and NW2 and also
between SE1 and SE2 but they still lack enough sensitivity
($\leq2\sigma$) in the PdBI-only map to be further substantiated at
this point. The brightest line in the spectra turns out to be NW2 and
C, whereas the other three features appear to be rather weak with a
signal-to-noise ratio at line peak of $\sim 4$
(Fig.~\ref{pdbi}). However, NW2 appears to be quite weak in the
integrated map which might be traced back to its weak line width of
only $\sim$50\kms. The integrated map was produced by averaging the
channel maps over a total velocity width of $\sim$600\kms. A
comparison between the global fluxes derived from the 30\,m and the
PdBI data again shows that a large part of the - diffuse, extended -
flux (at least $\sim 70\%$ of the total flux) must be resolved by the
interferometer due to missing short spacings.  Adding up all the PdBI
spectra leads to a total molecular gas mass of $M_{\rm gas} \sim 2.5
\times 10^7$\,\msun. This is a tenth\footnote{Besides the decreased
flux measured with the PdBI also the line width is lower by
$\lesssim20$\% with respect to the 30m data.} of the total molecular
gas mass derived from the 30\,m spectra of $\sim 2.4 \times
10^8$\,\msun (Pott et al.\ 2004).

  \begin{figure}[!]
   \centering
    \resizebox{\hsize}{!}{\rotatebox{-90}{\includegraphics{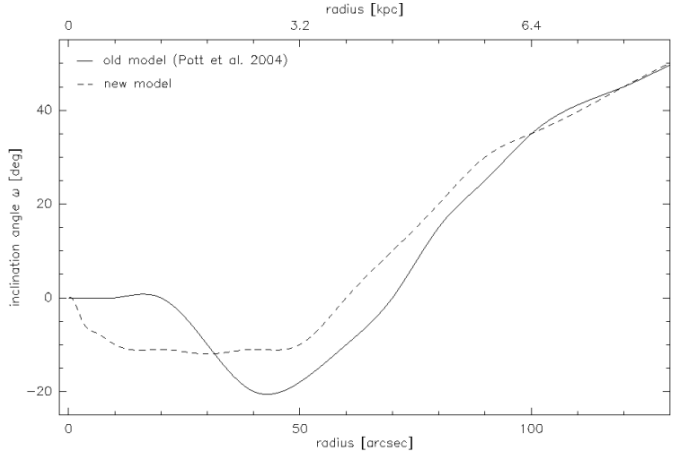}}}
      \caption{Tilt angles $\omega$(r). The old model of Pott et al.\
               (2004) is given as a solid line, while a dashed line is
               used for our new model.}
         \label{omega}
   \end{figure}

  \begin{figure}[!]
   \centering
    \resizebox{\hsize}{!}{\rotatebox{-90}{\includegraphics{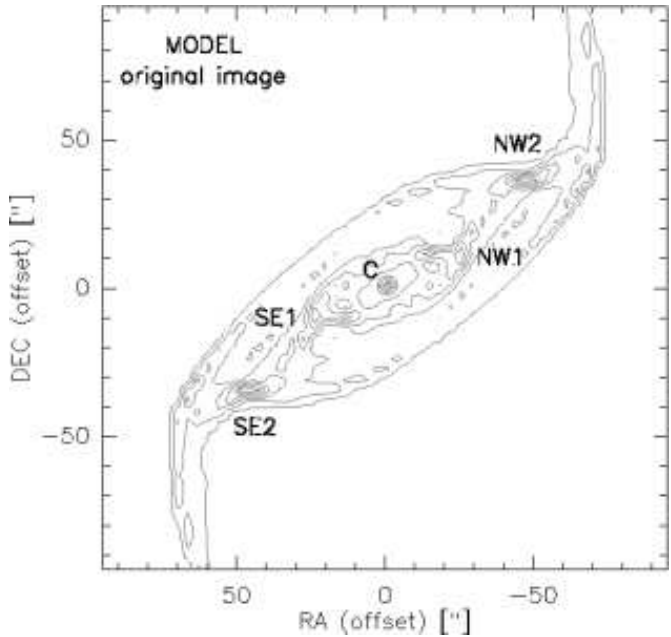}}}
      \caption{The integrated intensity map of our model is plotted
              here, before spatial filtering to match the response of
              the interferometer and single dish telescope.}
         \label{cube}
   \end{figure}

\subsection{PdBI+30m data: short spacing correction}
\label{datashsp}
To restore missing extended and diffuse emission, we combined the
CO(1--0) data from the IRAM 30\,m telescope with our interferometric
maps. We varied the weights attached to the 30m and PdBI data to find
the best compromise between good angular resolution and almost
complete restoration of the missing extended flux. In the end, a
factor of 4 was applied to the weights of the 30m data, leading to the
recovery of $\sim90$\% of the missing flux (see below) at a still
reasonable angular resolution of 7.8$\times$4.5$''$@20\degree. A
higher weighting factor would have increased the image restoration to
over 90\% but at the same time also decreased the resolution to
$\sim10''$, while a lower weighting factor would have had a higher
angular resolution but a worse image restoration due to a lower
signal-to-noise ratio. The combined maps of 30m and PdBI data have
finally been cleaned using standard program packages provided by the
GILDAS software for mosaic data. The results are presented in
Fig.~\ref{pdbicen-shsp} \& \ref{data-pdbi30m-r4} for the single field
and mosaic maps. In Fig.~\ref{velo-pdbi-pdbi30m}, we compare the
position-velocity diagrams along the axis used in
Fig.~\ref{pdbi-co10-cen-velo} for the PdBI-only data and the PdBI+30m
data. The asymmetry discussed in Section~\ref{line} is still present
even if slightly weakened and hence cannot be totally traced back to
missing short spacings. This indicates that the asymmetry seems to be
an intrinsic property of NGC~3718. A first comparison between the
PdBI-only and the PdBI+30\,m maps shows in the mosaic maps that the
short spacing correction strengthens the inner components (NW1, C and
SE1) that appear to contain most of the extended and diffuse
emission. If we now derive the total molecular gas mass, we obtain
$\sim$2.1$\times 10^8$\,\msun, which is $\sim$90\% of the value found
for the 30\,m-only data.  Besides the five main emission peaks seen in
Fig.~\ref{pdbi}, further emission between NW1 and NW2 and SE1 and SE2,
as already indicated in the PdBI-only map (Fig.~\ref{pdbi}), is
substantiated in the PdBI+30m map. Their gas masses are estimated to
$\sim 0.2 \times 10^8$\,\msun, respectively. Globally, the western
part of the CO(1--0) emission appears to contain more gas ($M_{\rm
gas} \simeq 10^8$\,\msun without C) than the eastern one ($M_{\rm gas}
\simeq 0.5 \times 10^8$\,\msun without C). The CO(1--0) emission
cannot be associated with a single straight line but rather delineates
an ``S-like'' shape.

The next sections will deal with the modeling of the complex gas
distribution in NGC~3718 for a better understanding and a more
quantitative interpretation. We will also examine in detail the
effects of missing short spacings on our data by applying our
procedure to a model and comparing the results thus obtained with the
observations.

\section{Modeling the CO distribution}
\label{sim}

The special characteristics of the H\,I data for NGC~3718 led to the
conclusion (Schwarz 1985) that the atomic gas is warped at radii
$\gtrsim 8$\,kpc in a configuration similar to that of the dust lane.
IRAM 30\,m observations of the molecular gas by Pott et al.\ (2004)
showed this warp extending down to smaller radii of $\sim
1$\,kpc. Previous modelling based on the Pott et al. CO and the H\,I
data can hence be fine-tuned with the new information gained from our
significantly increased spatial resolution-- $\sim 4-7''$ for our
PdBI-only map vs. 21$''$ for the Pott et al. (2004) 30\,m map.

\subsection{The model}
Our simulations of the mosaic maps are based on a tilted ring model
similar to that of Pott et al.\ (2004); that paper presents more
modelling details. Since the Pott et al. (2004) model of NGC~3718
reproduces the 30\,m observations with only 3 main maxima, we had to
modify the parameters to adapt their model to our PdBI results.  The
revised best-fit parameters were obtained by starting with the model
of Pott et al.\ (2004), propagating it through the PdBI+30m response
function (described in section\,\ref{shsp}), comparing the result with
the observed data (PdBI+30m), and finally tuning the parameters until
the match is optimal. In the new best-fit model, we mainly use
different tilt angles $\omega(r)$ (of a given ring relative to the
assumed plane of the innermost reference disk) for the inner radii
($\leq 50''$). Our inner rings are less warped, i.e., the tilt angles
are smaller for radii from 40$''$ to $60''$ than in the previous
model, but they remain constant down to radii of $\sim 1''$ (i.e., in
our model, the warp continues down to a radius of $\sim 1''$; see
Fig.~\ref{omega}).  Furthermore, we use a different intensity profile
from Pott et al. (2004). The earlier work assumed an integrated line
flux distribution for the rings that is very bright at the center but
then decreases rapidly, while the intensities in our model show a
slower decrease with radius. We also adopt a different inclination
(70\degree instead of 60\degree) and position angle (-60\degree
instead of -85\degree) for the galactic disc in space. The rotation
curve, however, remains the same for the new model. The integrated map
of our model is shown in Fig.~\ref{cube} (the model has not yet been
filtered through the interferometer or single dish response
function). While the central (C) and the two outer (SE2, NW2)
components are quite compact and bright, the two inner features (SE1,
NW1) are rather weak and relatively extended on $\sim 10''$ scales.

  \begin{figure*}[!]
   \centering
    \resizebox{17cm}{!}{\rotatebox{-90}{\includegraphics{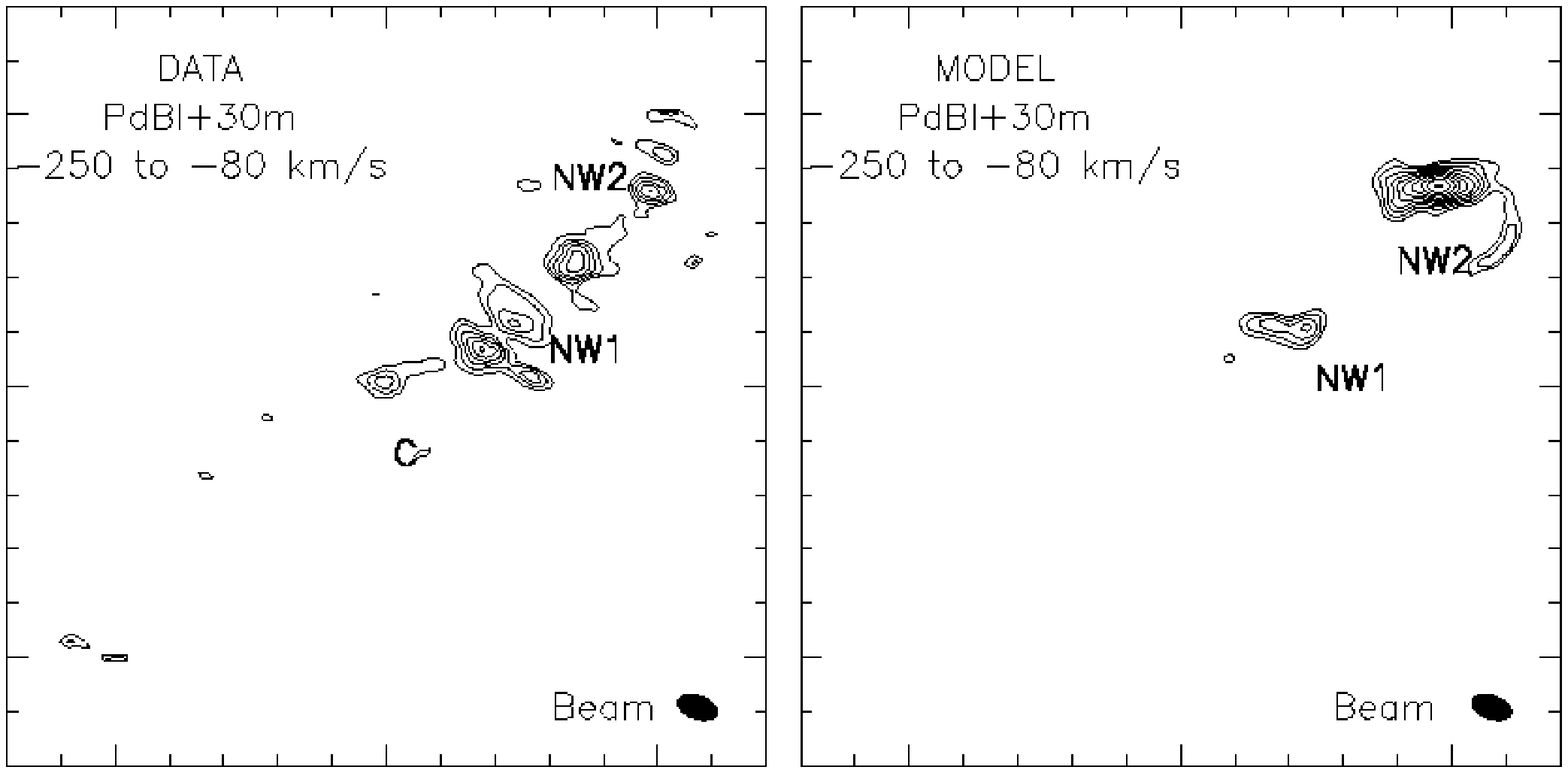}}}
    \resizebox{17cm}{!}{\rotatebox{-90}{\includegraphics{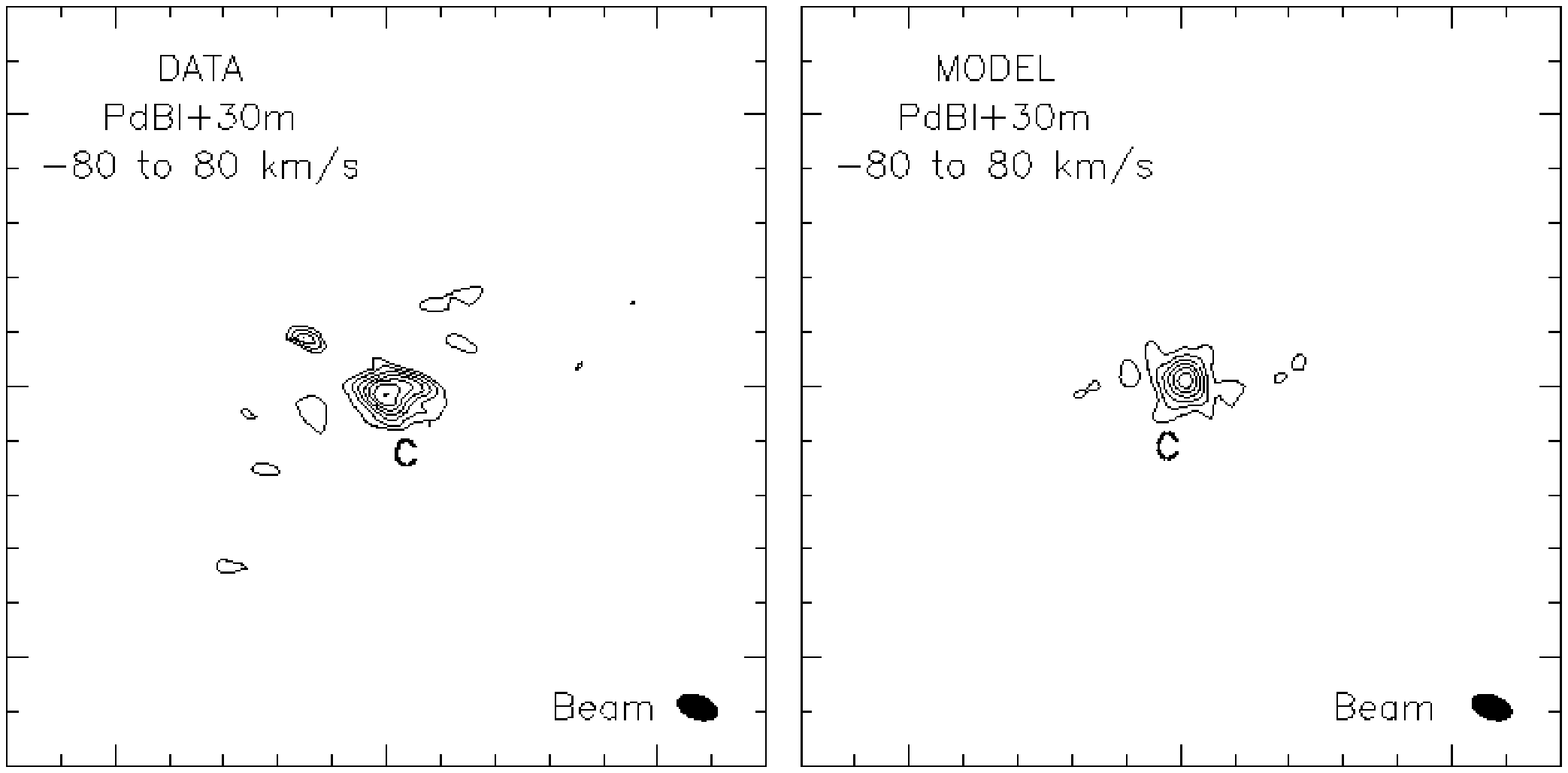}}}
    \resizebox{17cm}{!}{\rotatebox{-90}{\includegraphics{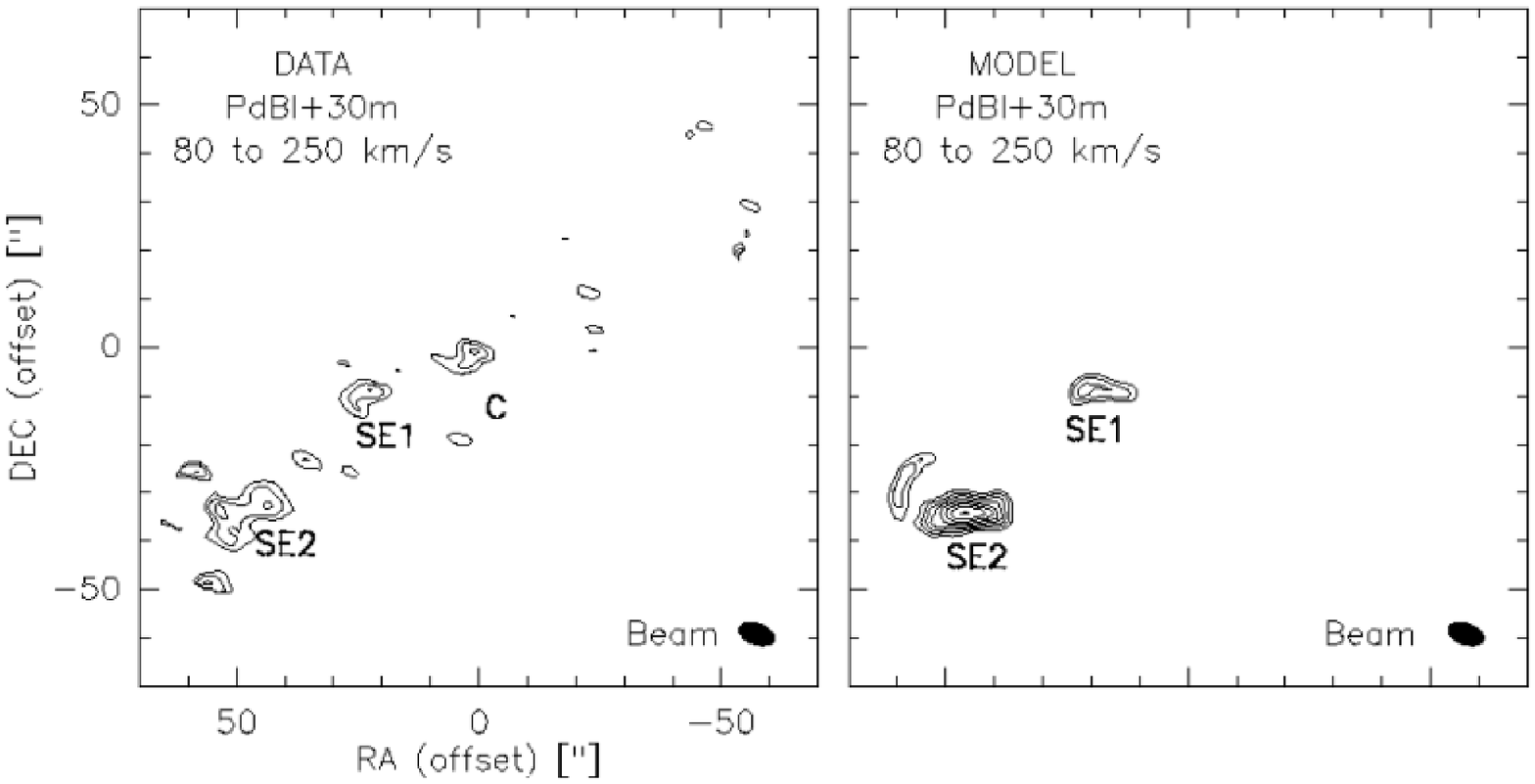}}}
      \caption{Model vs. Observation. Integrated maps of the measured
	({\it left}) and simulated ({\it right}) PdBI+30\,m data
	restricted to the following velocity ranges: -250 to -80\kms
	({\it upper panel}), -80 to 80\kms ({\it middle panel}), and
	80 to 225\kms ({\it lower panel}). Contour levels are
	$3\sigma=$1.2${\rm Jy\,beam^{-1}\,km\,s^{-1}}$ in steps
	of $1\sigma$.}
         \label{shsp-vel}
   \end{figure*}

  \begin{figure*}[!]
   \centering
    \resizebox{18cm}{!}{\rotatebox{-90}{\includegraphics{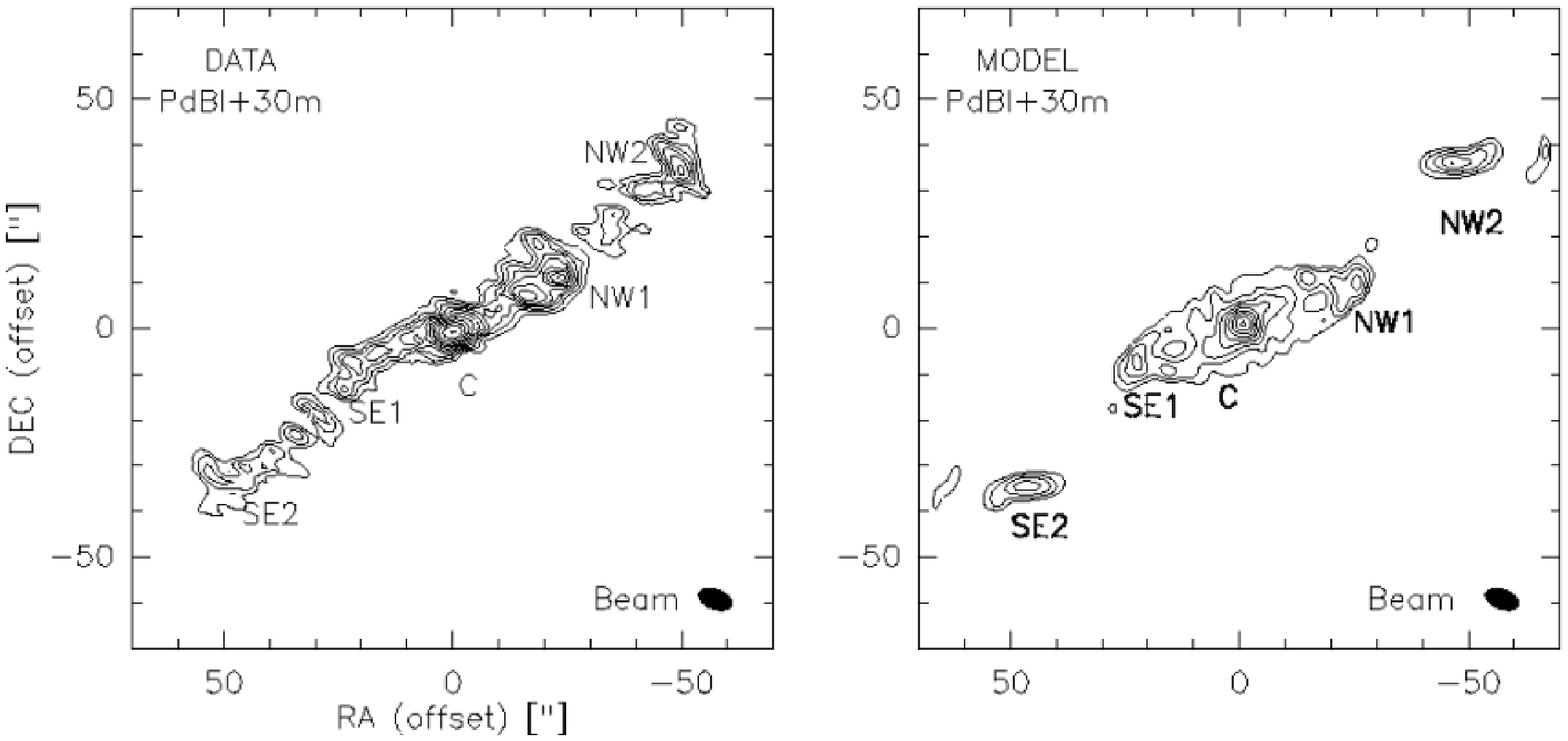}}}\\[0.4cm]
    \hspace*{-0.2cm}\resizebox{16cm}{!}{\rotatebox{-90}{\includegraphics{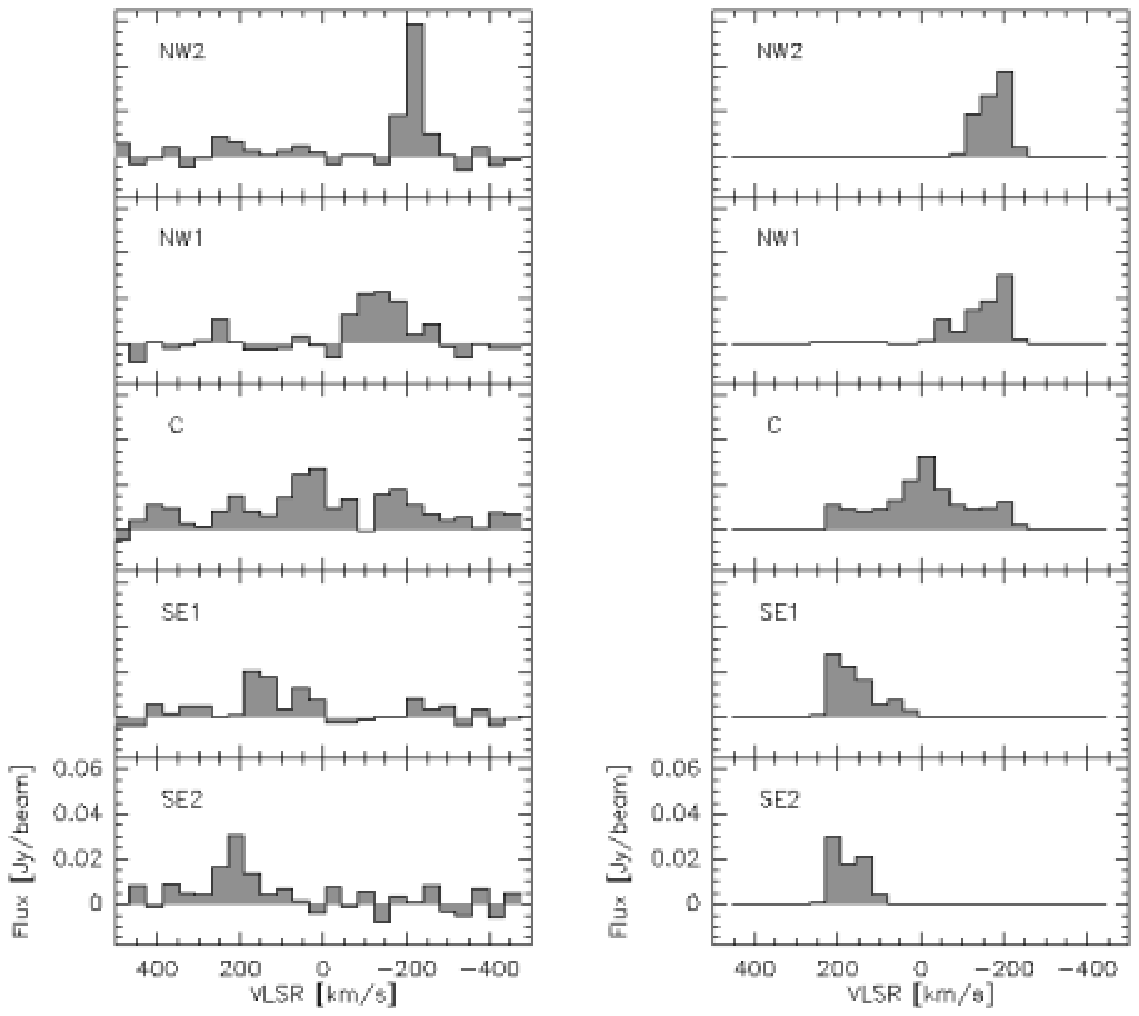}}}
      \caption{Upper panels: the short-spacing corrected images of the
      observed data ({\it left}) and of the model ({\it right})
      data. Lower panels: 30\,m spectra at positions of the CO
      peaks. Contour levels are from $2\sigma=$1.0 to
      6.0\,Jy\,beam$^{-1}$\,\kms in steps of $1\sigma$}.
         \label{mo-da-shsp-r4}
   \end{figure*}

  \begin{figure*}[!]
   \centering
    \resizebox{\hsize}{!}{\rotatebox{-90}{\includegraphics{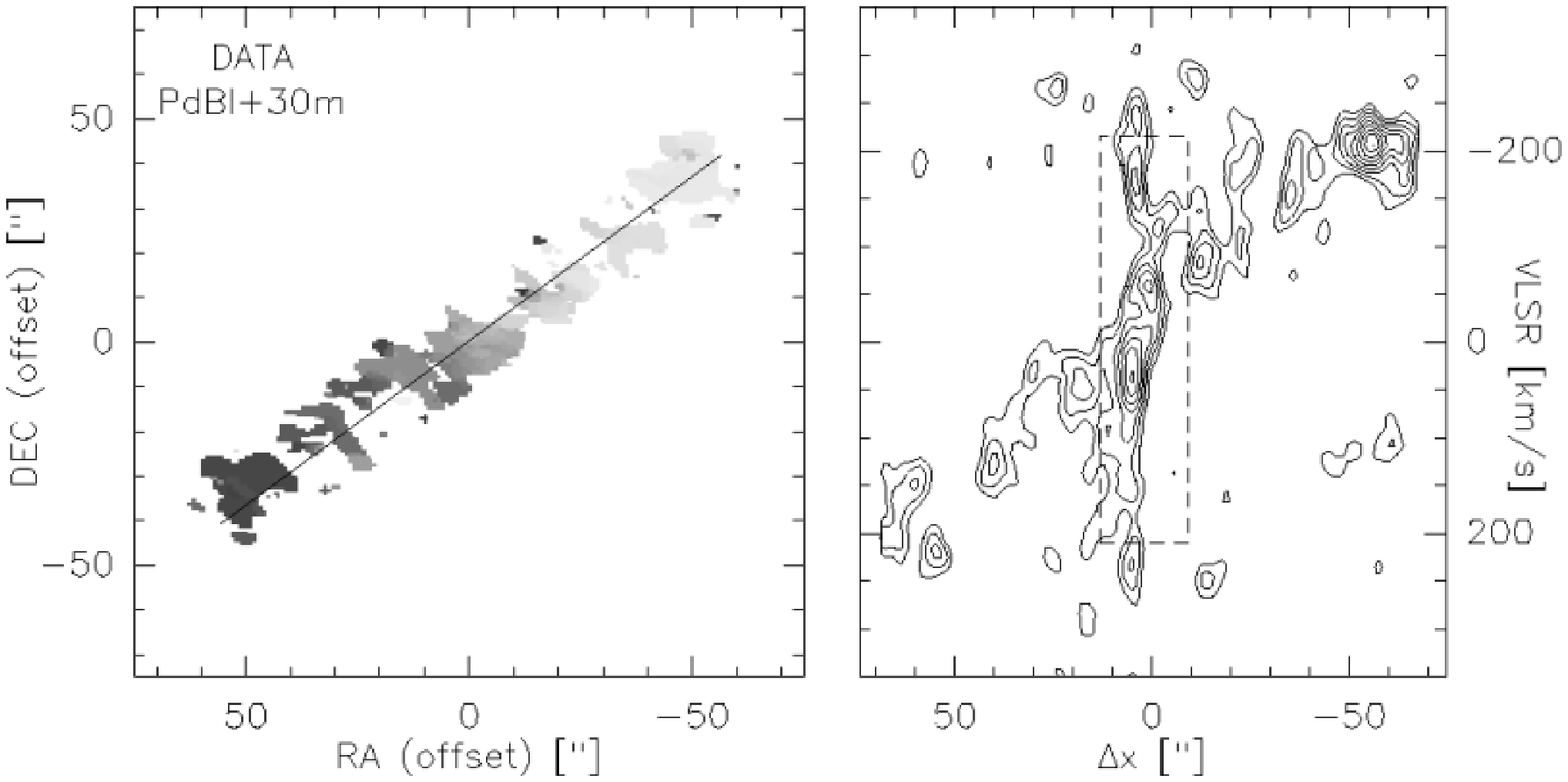}}}
    \resizebox{\hsize}{!}{\rotatebox{-90}{\includegraphics{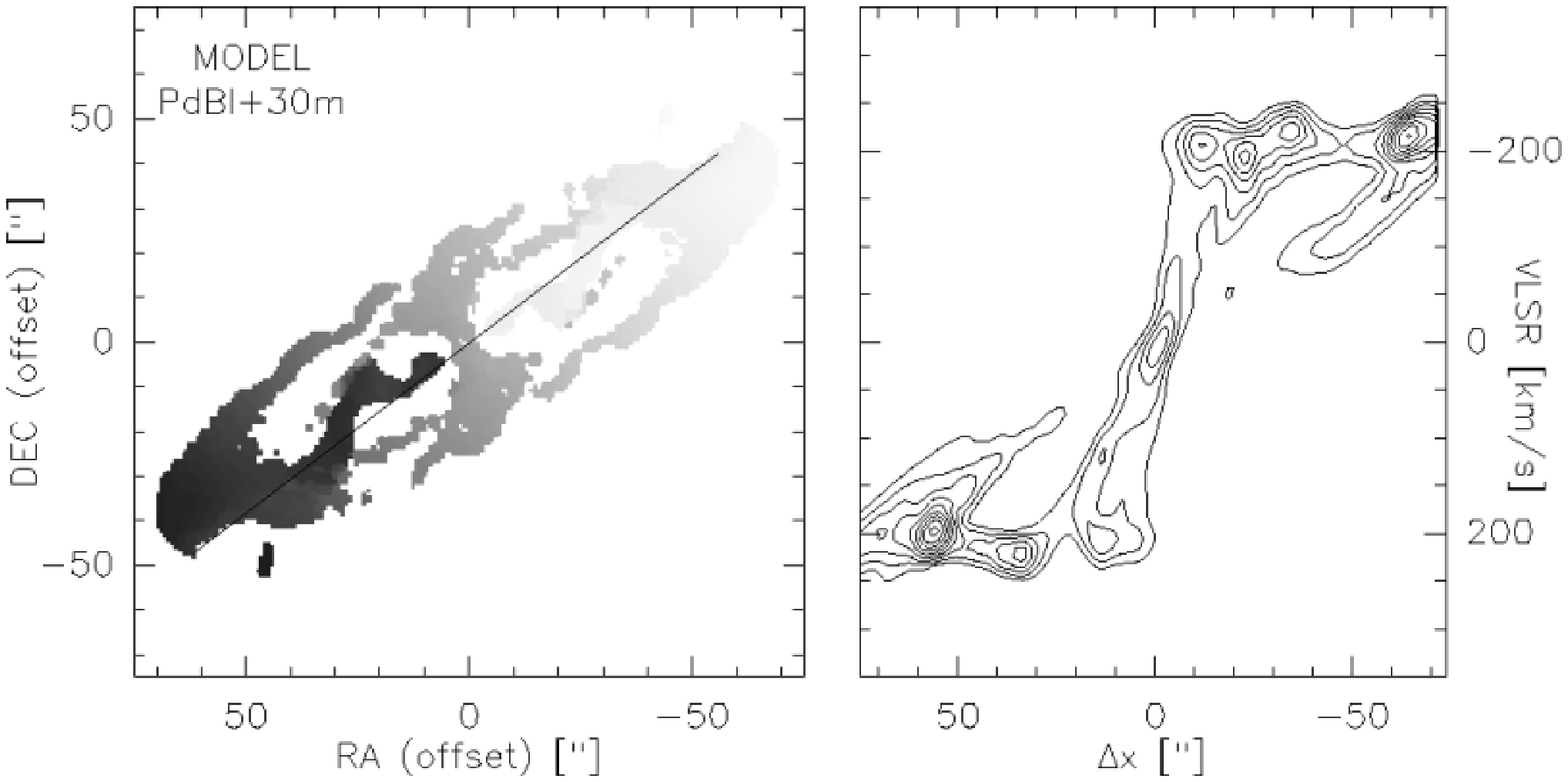}}}
      \caption{Upper panels: observed data, with first order moment
              maps ({\it left}) and position velocity cuts ({\it
              right}) taken along the axis indicated in the first
              order moment map (going through the AGN). Lower panels:
              same as above for the model. The area within the dashed
              box is identical with the map shown in the right panel
              of Fig.~\ref{velo-pdbi-pdbi30m}. Contour levels are in
              steps of 10\% from the peak starting at 20\%. The
              position velocity maps were Hanning smoothed. }
         \label{mosaic-velo}
   \end{figure*}

\subsection{Simulating the short-spacing correction}
\label{shsp}

To allow for a direct comparison between the derived model and the CO
data (PdBI-only, PdBI+30m, 30m-only), we simulated the effect of the
different observations on the model. To determine how our model galaxy
would look when observed with the interferometer (before and after
short-spacing correction), we first produced the simulated PdBI map
with a program that computes an interferometric image (upper panel of
Fig.~\ref{shspall}) from a given $uv$ coverage (taken here from the
observations) and the given three-dimensional intensity distribution
(taken here from our best-fit model shown in Fig.~\ref{cube}). Only
the most compact features are visible in the PdBI-only map while the
extended emission is resolved, resulting in a loss of around 60\% of
the original flux. These compact features correspond to the compact
orbital crowding points SE1, SE2, NW1 and NW2 ( Fig.~\ref{cube}). The
remaining emission from the tilted rings appears to be extended and
mainly concentrated around SE1 and NW1. It cannot be seen with the
PdBI alone due to spatial filtering, meaning that only the five
compact crowding points remain in the PdBI-only map. The extended
emission around SE1 and NW1 is thus heavily resolved (compare
Fig.~\ref{cube}). The best-fit model was also smoothed to the beam
($\sim 21''$) of the 30\,m telescope and the spectra were taken at the
same positions as for the observations. Accounting for normalisation,
these data were transformed into the 30\,m file that would have been
produced if such a distribution had been observed with the IRAM 30\,m
telescope. This file is thus suitable for comparison with the measured
30\,m data (see lower panel in Fig.~\ref{shspall}). We indeed find
only three main maxima, a merging of the inner two (SE1,NW1) with the
outer two maxima (SE2,NW2) and an S-line morphology.
 
As a final test, we combined the simulated 30\,m and PdBI maps. All
three simulated maps are plotted in Fig.~\ref{shspall} ({\it grey
scale}). The loss of flux due to the reduced sensitivity to diffuse
extended emission in the interferometer-only data is indeed reproduced
by the model. A comparison of the simulated peak fluxes from the PdBI,
the PdBI+30\,m, and the 30\,m data with each other suggests that the
PdBI has lost a factor $\sim 2-3$ of the peak line flux, consistent
with what is observed. The short-spacing correction not only
reproduces the correct fluxes but also the correct distribution
without any artefact or inconsistency in excess of 1\% for the model.

\subsection{Model vs. Data}
\label{movsda}

We have carried out the same steps to combine the observed PdBI and
30\,m data (section~\ref{datashsp}). The results are plotted in
Fig.~\ref{shspall} as contours over the simulated maps. We find a very
similar behavior in the observed and model maps. Besides the inner two
maxima which are more easily visible, the flux level also increases.
Thus, a better distinction between signal and noise is possible. As
the two eastern components are less dominant (by a factor 2) than the
western components in the integrated map over the whole velocity
range, we show separate plots for each component integrated over the
appropriate velocity range (see Fig.~\ref{shsp-vel}). Such an
asymmetry of the gas distribution has already been found in many other
galaxies as well (e.g., Richter \& Sancisi 1994) and cannot be
explained by our symmetric tilted ring model. It might be an
indication for a tidal interaction with the companion galaxy which
would result in an asymmetric gas distribution. Thus, our warp model
could presumably be improved by adding asymmetries to account for the
putative interaction.

The maps of the data ({\it left side}) and the model ({\it right
side}) are separately shown in Fig.~\ref{mo-da-shsp-r4}, along with
the good agreement between the model and the data in terms of spectra
taken at each of the five components. A somewhat weaker eastern
component is also visible in the model. This is due to the asymmetric
sampling of the 30\,m-spectra with respect to the CO
centroid. Simulated and observed line widths are consistent within the
noise.  In Fig.~\ref{mosaic-velo}, we have derived first order moment
maps with a $\sim3\sigma$ clipping over a velocity range of
$\sim-300$\kms to $+300$\kms and taken position-velocity cuts along
the major axis of the galaxy for the observed and simulated
data. Compared to the pv-diagram of a standard gas disk, a more
complex structure is seen in the pv-diagrams of the observations as
well as of the simulation due to orbital crowding effects. As might be
expected for a rapidly rotating nuclear disk with a size of $\sim
1.5\,{\rm kpc}$, a steep linear rise is indicated in the
position-velocity diagram.  This feature agrees well in orientation
and size with the central field velocity component visible in the
CO(1--0) map of Fig.~\ref{pdbi-co10-cen-velo}. We again find that this
central feature in the position-velocity map extends only over a range
of $\pm 10''$ in Fig.~\ref{mosaic-velo}, which is smaller than the
interval derived by Pott et al.\ (2004). The difference from the
position velocity diagram in Pott et al. (2004) has a logical and
simple explanation. In their maps, \{SE1, SE2\} and \{NW1, NW2\} merge
(in the 30\,m beam) into almost single SE and NW features,
respectively, and thus all lie on the major axis. In our
interferometric map, SE1 and NW1 are no longer aligned with this major
axis and thus appear only weakly in the position-velocity diagram of
Fig.~\ref{mosaic-velo}. The high velocity gas returns if the major
axis is rotated by about $\sim$10\degree. We thus have three different
velocity regimes. The first ranges from $\pm 10''$ with $\Delta v \sim
\pm \sim 150$\,\kms and can be identified with the gas occupying the
inner rings. The second region extends to $\pm 30''$ with $\Delta v
\sim \pm 220$\,\kms and is caused by intermediate rings. The third
extends to $\pm 70''$ with $\Delta v = \pm 220$\,\kms and originates
from the outer rings.

\section{Discussion \& Conclusions}

The paper presents an analysis based on arcsec angular resolution PdBI
observations of the gas distribution in NGC~3718, complementing
previous studies conducted by Pott et al.\ (2004: CO on $\geq 20''$
scales) and Schwarz (1985: H\,I on $\geq 30''$ scales). After making a
short-spacing correction using the IRAM 30\,m data (Pott et al.\
2004), we have modelled and interpreted the mosaic observations of the
total emission as well as single field observations of the central
kpc. Three compact and three extended components were found in the
mosaic map of the CO(1--0) emission.  At the higher angular resolution
of the PdBI, the gas structure turns out to be more complex than
previously appreciated. To explain the detected molecular gas
distribution, we were forced to modify the Pott et al. tilted ring
model slightly. The revised model now not only reproduces well the
PdBI+30\,m observations, but due to the higher resolution of the
observations, also places much tighter constraints on the warp.  The
disk is most likely already warped on arcsec scales ($\sim
4''\equiv250$\,pc) and is evident within a radius of 20$''$ (the
S-like shape between SE1, C, and NW1) even if the warp is less
dominant within 50$''$ relative to the original Pott et al. model. The
five symmetric CO(1--0) peaks (SE2, SE1, C, NW1, and NW2) detected by
the PdBI can be explained by orbit crowding effects. The two outer
maxima merge into one feature on each side of the center in the beam
of the IRAM 30\,m telescope and thus at this resolution appear as
single components. Besides the gas distribution, the gas kinematics
can be almost completely explained by the model. The position-velocity
diagrams of both data sets - mosaic and central pointing - unveil
rapidly rotating nuclear gas within a radius of $\sim 700$\,pc, as
seen in the simulated data. The high velocities in the center are
produced by the inner warped rings. Thus, the dynamics of the
molecular gas can be traced back to a continuous warped disk with
concentric orbits.

As in Pott et al.\ (2004), our model does not require a bar to explain
the observed gas properties down to $\sim 250$\,pc. However a warp
alone is not sufficient to account for all data, it must be combined
with asymmetry in gas density between the eastern and western side.
This asymmetry is logically coming from the same mechanism producing
the warp: i.e. an interaction with the companion galaxy.  The
asymmetric gas distribution within the central 250\,pc provides the
possibility for an accretion of gas onto the central engine. However,
the transport of gas to the black hole still raises many questions and
problems. Even in the Milky Way, where a warped circumnuclear
molecular torus (Guesten et al.\ 1987) provides a reservoir for
accretion at a distance of only $\sim 1\,{\rm pc}$ from the nucleus,
the situation is not yet understood.  In this context, the mechanisms
of gas fueling to the inner few pc/subpc in active nuclei appear even
more complex. Besides the comprehension of our own Galactic center,
only detailed studies of further nuclei with the highest angular
resolution can provide answers to the wealth of open questions.

\begin{acknowledgements}
The research presented in this paper has been financially supported in
part by the SFB 494. Stephane Leon is partially supported by DGI Grant
AYA 2002-03338 and Junta de Andaluc\'ia. We thank the IRAM staff from
the Plateau de Bure and Grenoble for conducting the observations and
helping with the data reduction. We are grateful to the referee
Dr. Michel Guelin for his useful comments.

\end{acknowledgements}

\end{document}